\documentclass[journal=jacsat,manuscript=article]{achemso}
\usepackage{chemformula} 
\usepackage[T1]{fontenc} 

\usepackage{float}
\usepackage{graphicx}
\usepackage{pdfpages}
\usepackage{dcolumn}
\usepackage{bm}
\usepackage{mathrsfs,amssymb,amsmath}
\usepackage{ulem}
\usepackage{caption}
\usepackage[colorlinks=true,linkcolor=black,citecolor=black,anchorcolor=black,urlcolor=black]{hyperref}
\usepackage{booktabs}

\usepackage{setspace}

\author{Yan Yin}
\affiliation[NUAA]
{Institute for Frontier Science, Institute of Nanoscience, State Key Lab of Mechanics and Control of Mechanical Structures \& Key Lab for Intelligent Nano Materials and Devices of Ministry of Education \& College of Aerospace Engineering, Nanjing University of Aeronautics and Astronautics (NUAA), Nanjing 210016, China}
\author{Min Yi}
\email{yimin@nuaa.edu.cn}
\affiliation[NUAA]
{Institute for Frontier Science, Institute of Nanoscience, State Key Lab of Mechanics and Control of Mechanical Structures \& Key Lab for Intelligent Nano Materials and Devices of Ministry of Education \& College of Aerospace Engineering, Nanjing University of Aeronautics and Astronautics (NUAA), Nanjing 210016, China}
\alsoaffiliation[NU]
{National Laboratory of Solid State Microstructures, Nanjing University, Nanjing 210093, China}
\author{Wanlin Guo}
\email{wlguo@nuaa.edu.cn}
\affiliation[NUAA]
{Institute for Frontier Science, Institute of Nanoscience, State Key Lab of Mechanics and Control of Mechanical Structures \& Key Lab for Intelligent Nano Materials and Devices of Ministry of Education \& College of Aerospace Engineering, Nanjing University of Aeronautics and Astronautics (NUAA), Nanjing 210016, China}

\title[AnTitle]{High and anomalous thermal conductivity in monolayer MSi$_2$Z$_4$ semiconductors}

\abbreviations{IR,NMR,UV}

\begin{document}    
\begin{abstract}
The lattice thermal conductivity ($\kappa$) of newly synthesized two-dimensional (2D) MoSi$_2$N$_4$ family and its associated abnormality are anatomized by \textit{ab initio} phonon Boltzmann transport calculations. $\kappa$ of MoSi$_2$N$_4$ and WSi$_2$N$_4$ is found over 400~Wm$^{-1}$K$^{-1}$ at 300~K. $\kappa$ of MoSi$_2$Z$_4$ (Z\,=\,N,\,P,\,As) obeys Slack's rule of thumb, decreasing by one order of magnitude from Z\,=\,N to Z\,=\,As with 46~Wm$^{-1}$K$^{-1}$. However, in MSi$_2$N$_4$ (M\,=\,Mo,\,Cr,\,W,\,Ti,\,Zr,\,Hf), the variation of $\kappa$ with respect to M is anomalous, i.e. deviating from Slack's classic rule. For M in the same group, $\kappa$ of MSi$_2$N$_4$ is insensitive to the average atomic mass, Debye temperature, phonon group velocity, and bond strength owing to the similar phonon structure and scattering rates. MSi$_2$N$_4$ with heavy group-\uppercase\expandafter{\romannumeral6}B M even possesses a three to four times higher $\kappa$ than that with light group-\uppercase\expandafter{\romannumeral4}B M, due to its much stronger M-N and exterior Si-N bonds and thus one order of magnitude lower phonon scattering rates. Nevertheless, this abnormality could be traced to an interplay of certain basic vibrational properties including the bunching strength and flatness of acoustic branches and their nearby optical branches, which lie outside of the conventional guidelines by Slack. This work predicts high $\kappa$ of 2D MSi$_2$Z$_4$ for thermal management and provides microscopic insight into deciphering the anomalous $\kappa$ of layered 2D structures.
\end{abstract}

\textbf{Keywords}: 2D MoSi$_2$N$_4$ family, Lattice thermal conductivity, Phonon vibration and scattering, \textit{ab initio} calculations, Phonon Boltzmann transport equation

\textcolor{blue}{The work was submitted to ACS Applied Materials \& Interfaces and was accepted for publication in Sep 6, 2021.}

\section{1. INTRODUCTION}
Two-dimensional (2D) semiconductors, such as black phosphorene (BP)~\citep{2014-BlackP-electricFieldEffect-apl}, semiconducting transition metal dichalcogenides (TMDs)~\cite{2014-TMDs-APL-guxiaokun} and group-\uppercase\expandafter{\romannumeral4}A and -\uppercase\expandafter{\romannumeral6}A compounds~\cite{2016GroupIV-VI-A, 2018GroupIVSe}, have aroused considerable interests in various nanoelectronic devices, due to their outstanding electrical properties, flexibility, and controllable bandgap.
The interesting thermal transport properties in 2D semiconductors induced by quantum confinement effect provides new directions for thermal energy control and management, such as interface thermal resistance~\cite{2015-MoS2-Au-InterfaceResistance-ACSAMI, 2015-MoS2-MoSe2-Interfacial-ACSAMI}, thermoelectric~\cite{2014-SnSe-Thermoelectric, 2016-BlackP-Thermoelectric} and thermal transistor devices~\cite{2014-BlackP-fieldEffectTransistors-NatureNanotech, 2017-TMD-InternationalJ.HeatMassTransfer}. On the other hand, thermal dissipation is increasingly significant in integrated circuits since the power dissipation density and high-temperature hot spots cause the performance degradation as well as long-term reliability issues. 2D semiconductors with high thermal conductivity ($\kappa$) are critical for keeping nanoelectronic devices cool by removing the high-density dissipation. Up to now, compared with the high $\kappa$ of zero-bandgap graphene (3000--5000~Wm$^{-1}$K$^{-1}$)~\cite{2008Superior} and insulated hexagonal boron nitride (600~Wm$^{-1}$K$^{-1}$)~\cite{2019-BN-BP-BAs-BSb}, high $\kappa$ of 2D semiconductors is mainly in carbon- or boron-based materials~\cite{2018-C3N-nanoscale, 2018-C2N2-carbon, 2019-BC3-carbon, 2020-HYX-diamane}. 
Nevertheless, the exploration of new 2D semiconductors and their thermal properties never stops. 

Very recently, layered 2D MoSi$_2$N$_4$ and WSi$_2$N$_4$ have been synthesized successfully by introducing silicon to passivate the high-energy surfaces of non-layered molybdenum nitride during chemical vapor deposition growth~\cite{2020Chemical}. After the breakthrough discovery, the MA$_2$Z$_4$ family has been theoretically predicted~\cite{2021-MAZ-NatureCommunication}. The MA$_2$Z$_4$ phase contains early transition metal elements~(M), silicon or germanium~(A), and group-\uppercase\expandafter{\romannumeral5}A elements~(Z). In the large MA$_2$Z$_4$ family there exist affluent and interesting physical and chemical properties, for instances, piezoelectricity, photocatalysis, and strain-induced change of electrical properties~\cite{MORTAZAVI2021-NanoEnergy, PhysRevB.Valley-dependent, PRB.ValleyPseudospin, PhysRevB.strain-electrical, apl-electricalContact-MoSi2N4, 2021-MA2Z4-ElectronicMagneticCatalytic, 2021-MAZ-Nanoscale}. 
The monolayer MoSi$_2$N$_4$ shows high optical transmittance~(97.6$\%$), which is comparable to that of graphene~(97.7$\%$). Meanwhile, the excellent mechanical properties (Young's modulus of 491.4~$\pm$~139.1 GPa and breaking strength of 65.8~$\pm$~18.3 GPa) and the stability under ambient conditions laid out the promising usage of MoSi$_2$N$_4$ in the future~\cite{2020Chemical}. The high $\kappa$ of MoSi$_2$N$_4$ and WSi$_2$N$_4$ are appealing for the thermal management in semiconductor devices~\cite{MORTAZAVI2021-NanoEnergy, MoSiN-WSiN-DFT-k_2021}. Mortazavi and coworkers~\cite{MORTAZAVI2021-NanoEnergy} proposed that thermal conductivity of MA$_2$Z$_4$ is attributed to the core atomic mass, phonon group velocity and phonon-phonon scattering. Apart from these, many vital factors, such as average atomic mass, Debye temperature, phonon transport properties, and even the structure itself, which are critical for deciphering the underlying mechanism of $\kappa$ of MSi$_2$Z$_4$, still remain to be explored. 
 
In this paper, the effects of different M or Z atoms on structure and phononic nature of MSi$_2$Z$_4$ are discussed. Then, the thermal transport properties are revealed by solving phonon Boltzmann transport equation (BTE). 
The role of several possible factors (e.g. average atomic mass, Debye temperature, phonon group velocity, phonon-phonon scattering properties, bond strength, etc.) in determining $\kappa$ of MSi$_2$Z$_4$ is exhaustively examined. We find that MoSi$_2$N$_4$ and WSi$_2$N$_4$ possess a high $\kappa$ greater than 400~W$^{-1}$K$^{-1}$ at 300~K. 
In Slack's rule~\cite{1962-slack, 1973-Slack}, nonmetallic crystals with high thermal conductivity usually have four characters which are (1) low atomic mass, (2) strong bonding, (3) simple crystal structure, and (4) low anharmonicity.
However, the variation of $\kappa$ with respect to M atoms is anomalous and does not strictly obey Slack's classic rule. According to the rule of thumb by Slack, $\kappa$ decreases with increasing average atomic mass ($\overline{m}$) and decreasing Debye temperature ($\varTheta_\text{D}$). In contrast, $\kappa$ of MSi$_2$N$_4$ with larger $\overline{m}$ and smaller $\varTheta_\text{D}$ (M in group-\uppercase\expandafter{\romannumeral6}B) is even 3--4 times higher than that with smaller $\overline{m}$ and larger $\varTheta_\text{D}$ (M in group-\uppercase\expandafter{\romannumeral4}B). Moreover, $\kappa$ of MSi$_2$N$_4$ remains at the same level irrespective of $\overline{m}$ and $\varTheta_\text{D}$ when M atoms are in the same group. 
More microscopic details on deciphering the high and anomalous $\kappa$ of MSi$_2$Z$_4$ are also comprehensively examined.

\section{2. RESULTS AND DISCUSSION}
\subsection{2.1. Lattice structure and phonon band structure} \label{section1}
\begin{figure}[!t]
\centering
\includegraphics[width=11cm]{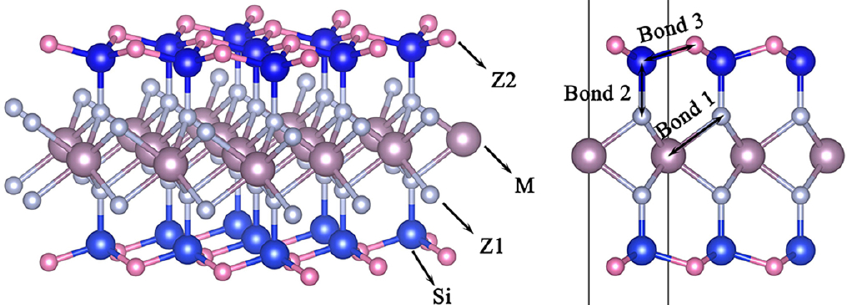}
\caption{Lattice structure of MSi$_2$Z$_4$. M is Mo, Cr, W, Ti, Zr or Hf atom, and Z is N, P, or As atom.}
\label{fig1}
\end{figure}

Herein, we investigate eight types of 2D MSi$_2$Z$_4$, including the experimentally synthsized MoSi$_2$N$_4$ and WSi$_2$N$_4$. 
The monolayer MSi$_2$Z$_4$ can be considered as a sandwich structure that one 2H MoS$_2$-type MZ$_2$ layer is inserted into an $\alpha$-InSe-type Si$_2$Z$_2$ layer. The perspective and side views of the structure are shown in Fig.~\ref{fig1}. There are four nonequivalent atoms in the unit cell, namely, M, Si, Z1 and Z2. The M represents transition metal elements (Mo,\,W,\,Cr,\,Ti,\,Zr,\,Hf) and the Z represents group-\uppercase\expandafter{\romannumeral5}A elements (N,~P,~As). Three different bonds formed by these atoms are denoted by Bond 1 (M-Z1 bond), Bond 2 (Si-Z1 bond) and Bond 3 (Si-Z2 bond). The structure parameters after full relaxation are listed in Table \ref{table1}. The lattice constant of MoSi$_2$N$_4$ is $a$\,=\,$b$\,=\,2.91~\AA, which is in line with the previous experimental results (2.909~\AA). For conveniently exploring the effect of different atoms on the properties of MSi$_2$Z$_4$, we divide the eight structures into two groups, i.e. MSi$_2$N$_4$ and MoSi$_2$Z$_4$.

As shown MoSi$_2$Z$_4$ in Table \ref{table1}, the lattice constants of MoSi$_2$P$_4$ and MoSi$_2$As$_4$ is 3.47 and 3.62~\AA, which are about 19.2$\%$ and 24.3$\%$ larger than that of MoSi$_2$N$_4$. Accordingly, the bonds are also elongated with the increasing period number of elements Z from N to P to As. The effective thickness ($h_\text{eff}$) along vertical direction is defined as the sum of distance between top and bottom atoms and vdW diameter of outmost Z2 atom. $h_\text{eff}$ of MoSi$_2$N$_4$, MoSi$_2$P$_4$ and MoSi$_2$As$_4$ are 10.01, 13.17 and 13.90~\AA, respectively. 
In contrary, for MSi$_2$N$_4$ all the lattice constant, bond length and effective thickness do not change remarkably with respect to M atoms.
With PBE calculation, the eight types of MSi$_2$Z$_4$ are all semiconductors with a band gap around of 0.49--2.08~eV, as shown in Fig.~S1. Apart from MoSi$_2$P$_4$ and MoSi$_2$As$_4$ with narrow direct band gap, other configurations are indirect semiconductors. The band gaps calculated by both GGA and HSE functionals are list in Table S1, indicating the semiconductor nature of MSi$_2$Z$_4$ family.
\begin{table*}[htbp]
\centering
\caption{Lattice parameters ($a=b$), effective thickness ($h_\text{eff}$), bond lengths, average mass ($\overline{m}$), group velocity ($\nu$) near $\Gamma$ point, and Debye temperature ($\varTheta_\text{D}$) of acoustic branches in 2D MSi$_2$Z$_4$.} 
\renewcommand\arraystretch{1.5} 
\setlength{\tabcolsep}{1.5mm}
\begin{tabular}{ccccccccccc}
\toprule
& $a=b$ &$h_\text{eff}$ &Bond~1 &Bond~2 &Bond~3 &$\overline{m}$ &$\nu_\text{ZA}$ &$\nu_\text{TA}$ &$\nu_\text{LA}$ &$\varTheta_\text{D}$ \\ 
&(\AA) &(\AA) &(\AA) &(\AA) &(\AA) &(amu) &(km/s) &(km/s) &(km/s) &(K)\\
\midrule
MoSi$_2$N$_4$ &2.91 &10.01 &2.10 &1.75 &1.76 &29.73 &4.10 &6.33 &10.30 &434.78\\ 
MoSi$_2$P$_4$ &3.47 &13.17 &2.46 &2.24 &2.25 &39.43 &4.03 &4.25 &6.04 &212.21 \\
MoSi$_2$As$_4$ &3.62 &13.94 &2.56 &2.36 &2.37 &64.5 &3.07 &3.09 &4.44 &144.92 \\
WSi$_2$N$_4$ &2.91 &10.01 &2.10 &1.75 &1.76 &42.29 &4.62 &5.45 &8.55 &336.28 \\
CrSi$_2$N$_4$ &2.84 &9.87 &2.00 &1.75 &1.73 &23.45 &3.88 &6.72 &11.20 &485.71\\
TiSi$_2$N$_4$ &2.93 &9.91 &2.07 &1.75 &1.77 &22.86 &3.92 &6.82 &10.90 &264.24\\ 
ZrSi$_2$N$_4$ &3.04 &10.05 &2.18 &1.75 &1.82 &29.06 &4.03 &5.77 &8.93 &208.94\\ 
HfSi$_2$N$_4$ &3.02 &10.00 &2.16 &1.75 &1.81 &41.52 &3.73 &4.83 &8.32 &168.00 \\
\bottomrule
\end{tabular}
\label{table1}
\end{table*}

\begin{figure*}[!t]
\centering
\includegraphics[width=16cm]{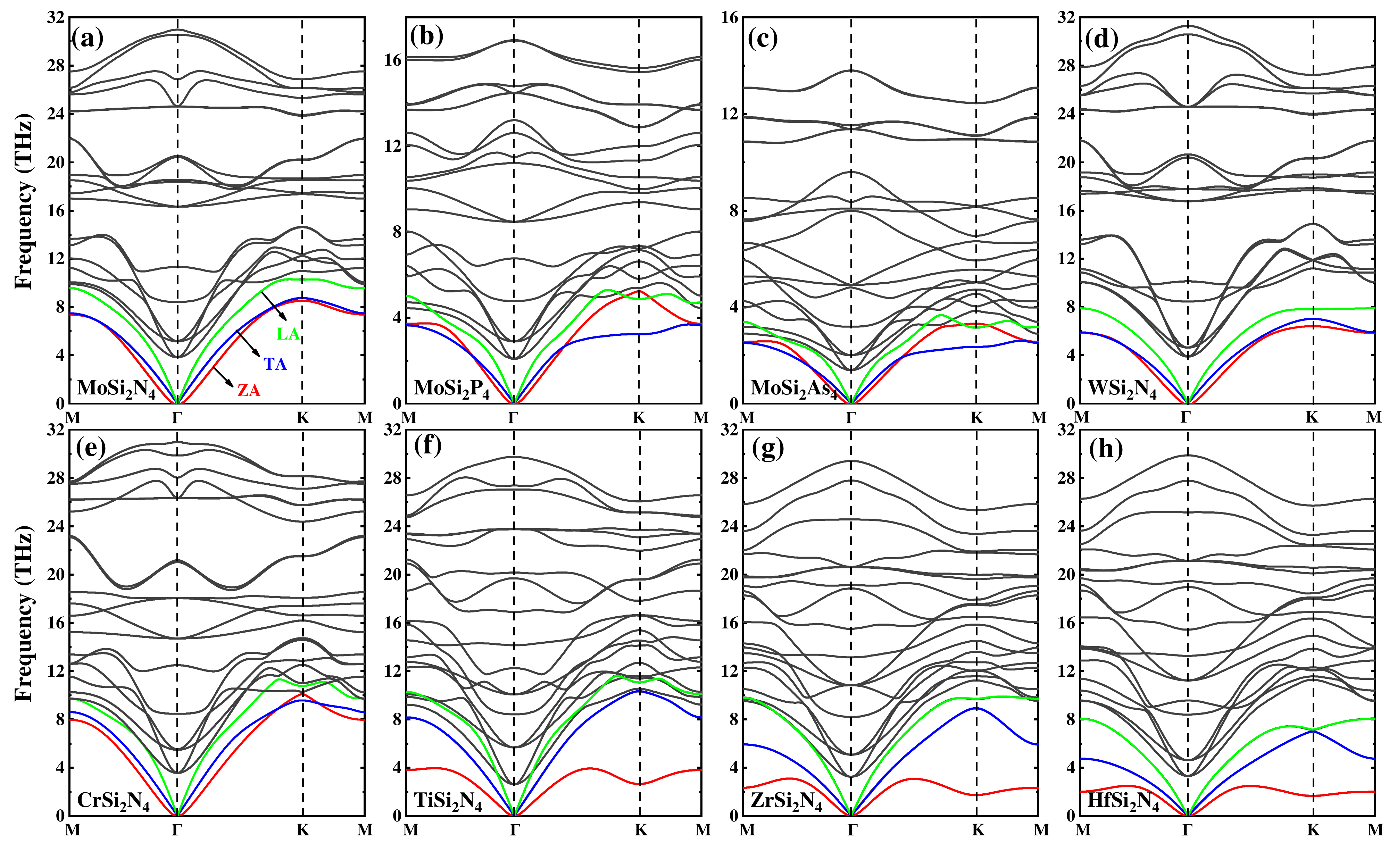}
\caption{Phonon dispersion spectrum of 2D MSi$_2$Z$_4$: (a)~MoSi$_2$N$_4$, (b)~MoSi$_2$P$_4$, (c)~MoSi$_2$As$_4$, (d)~WSi$_2$N$_4$, (e)~CrSi$_2$N$_4$, (f)~TiSi$_2$N$_4$, (g)~ZrSi$_2$N$_4$ and (h)~HfSi$_2$N$_4$.}
\label{fig2}
\end{figure*}

The phonon dispersion spectrums of MSi$_2$Z$_4$ are shown in Fig.~\ref{fig2}. There are 3 acoustic branches (including ZA, TA and LA branches) and 18 optical branches due to 7 atoms in the unit cell. No imaginary frequency of each branch indicates the dynamic stability. The phonon frequency range is around 0--30~THz in MSi$_2$N$_4$ (M\,=\,Mo,\,W,\,Cr,\,Ti,\,Zr,\,Hf), but it is only 0--17.02~THz in MoSi$_2$P$_4$ and 0--14~THz in MoSi$_2$As$_4$. Comparing the average mass ($\overline{m}$) of each structure in Table~\ref{table1}, one can find large $\overline{m}$ induces obvious softening of phonon branches.
In the phonon density of states (PhDOS) (Fig.~S2), heavy atoms occupy the low frequency region while light ones contribute to the high frequency region. 
Compared with N, P or As is heavier and thus PhDOS and the whole phonon spectrum of MoSi$_2$P$_4$ or MoSi$_2$As$_4$ notably shift to the lower frequency.
However, the case for MSi$_2$N$_4$ is entirely different. Firstly, the whole frequency range of phonon spectra of MSi$_2$N$_4$ remains almost the same in spite of different M atoms or $\overline{m}$. Secondly, heavier M or larger $\overline{m}$ prominently results in flatter lower-frequency acoustic branches, only when M in the same group. Thirdly, M in group-\uppercase\expandafter{\romannumeral4}B leads to much flatter acoustic branches (especially ZA branch) than M in group-\uppercase\expandafter{\romannumeral6}B. The flatter and lower-frequency ZA branch indicates that the out-of-plane vibration and interatomic force of MSi$_2$N$_4$ (M\,=\,Ti,\,Zr,\,Hf) along the $z$ direction are weaker.

\subsection{2.2. High $\kappa$ of MSi$_2$N$_4$ with Group-\uppercase\expandafter{\romannumeral6}B M} \label{section2}
\begin{figure*}[!t]
\centering
\includegraphics[width=14cm]{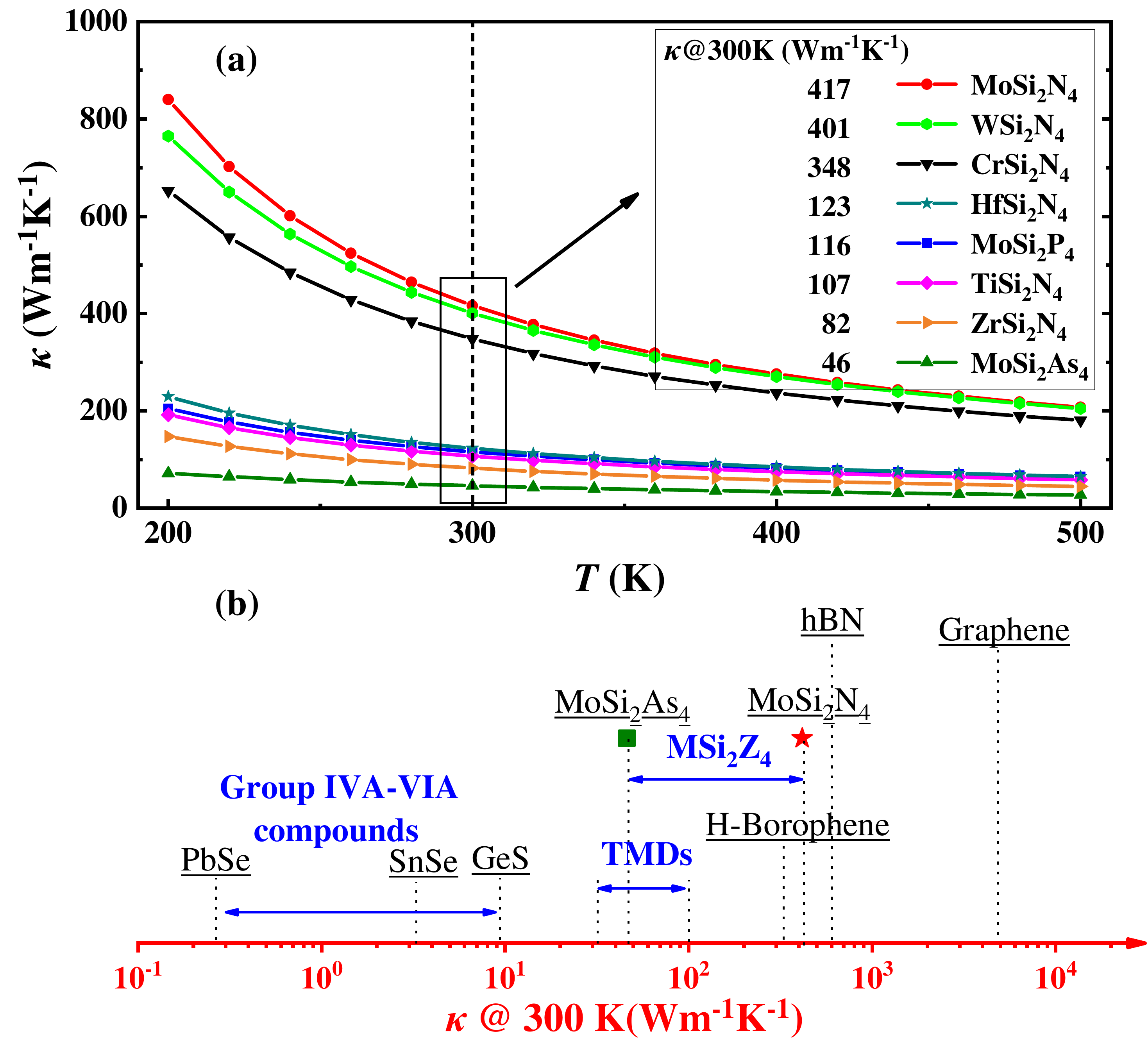}
\caption{(a) Thermal conductivity ($\kappa$) of MA$_2$Z$_4$ as the function of temperature (T), (b) the comparison of $\kappa$ in different 2D materials at 300 K, Ref.~\cite{2018GroupIVSe, 2008Superior, 2020BHBFBCl, 2014MoS2, 2014-TMD-GuXiaoKun, 2019-TMD, 2019-SnSe, 2008Graphene3000-APL, 2018MonolayerC3N, 2011BN600}.}
\label{fig3}
\end{figure*}

The calculated $\kappa$ of MSi$_2$Z$_4$ as a function of temperature ($T$) is presented in Fig.~\ref{fig3}\,(a). In the whole temperature range, $\kappa$ of MoSi$_2$N$_4$ is the highest and that of MoSi$_2$As$_4$ is the lowest. At room temperature (300~K), $\kappa$ of MoSi$_2$N$_4$ is up to 417~Wm$^{-1}$K$^{-1}$, which is higher than that of most other 2D materials, such as h-BAs (310~Wm$^{-1}$K$^{-1}$)~\cite{2021-hyx-biBAs-mtp}, hydrogenated borophene (368~Wm$^{-1}$K$^{-1}$)~\cite{2020BHBFBCl}, monolayer GaGeTe (58~Wm$^{-1}$K$^{-1}$)~\cite{2020-GaGeTe}, 2H transition metal dichalcogenides (e.g. WS$_2$ (142~Wm$^{-1}$K$^{-1}$), MoS$_2$ (23~Wm$^{-1}$K$^{-1}$), MoSe$_2$ (54~Wm$^{-1}$K$^{-1}$))~\cite{2014MoS2, 2014-TMD-GuXiaoKun, 2019-TMD}, and group-\uppercase\expandafter{\romannumeral4}A and -\uppercase\expandafter{\romannumeral6}A compounds (e.g. GeS (9.8~Wm$^{-1}$K$^{-1}$), PbSe (0.26~Wm$^{-1}$K$^{-1}$))~\cite{2018GroupIVSe, 2019-SnSe}. But it is lower than that of graphene (3000--5000~Wm$^{-1}$K$^{-1}$)~\cite{2008Superior, 2008Graphene3000-APL}, Carbon-based nitrides(820~Wm$^{-1}$K$^{-1}$)~\cite{2018MonolayerC3N} and hexagonal boron nitrogen (h-BN, 600~Wm$^{-1}$K$^{-1}$)~\cite{2011BN600}. Besides, the room temperature $\kappa$ of WSi$_2$N$_4$ and CrSi$_2$N$_4$ are 401 and 348~Wm$^{-1}$K$^{-1}$, respectively. As shown in Fig.~\ref{fig3}\,(b), the thermal conductive range of MSi$_2$Z$_4$ is wider than that of group-\uppercase\expandafter{\romannumeral4}A and -\uppercase\expandafter{\romannumeral6}A compounds or TMDs. And such high $\kappa$ of group-\uppercase\expandafter{\romannumeral6}B MSi$_2$N$_4$ is promising for heat conductors and thermal management in semiconductor devices. The underlying microscopic mechanism for the high $\kappa$ in MSi$_2$N$_4$ with group-\uppercase\expandafter{\romannumeral6}B M will be anatomized in Section~2.4. 

\subsection{2.3. Large reduction of $\kappa$ in MoSi$_2$Z$_4$} \label{section3}
At 300~K, $\kappa$ of MoSi$_2$Z$_4$ remarkably decreases from 417~Wm$^{-1}$K$^{-1}$ of MoSi$_2$N$_4$ to 116~Wm$^{-1}$K$^{-1}$ of MoSi$_2$P$_4$ to 46~Wm$^{-1}$K$^{-1}$ of MoSi$_2$As$_4$ (Fig.~\ref{fig3}). In fact, h-BZ (Z\,=\,N,\,P,\,As)~\cite{2019-BN-BP-BAs-BSb} and graphene-like BZ (Z\,=\,H,\,F,\,Cl)~\cite{2020BHBFBCl} also show the similar phenomenon. But their reduction in $\kappa$ with respect to Z is much smaller than that in MoSi$_2$Z$_4$. For instance, $\kappa$ of h-BAs at 300~K is 40\% of that of h-BN, whereas $\kappa$ of MoSi$_2$As$_4$ is only 11\% of that of MoSi$_2$N$_4$. The large reduction of $\kappa$ of MoSi$_2$Z$_4$ could be ascribed to the following factors.

\begin{figure}[!t]
\centering
\includegraphics[width=11cm]{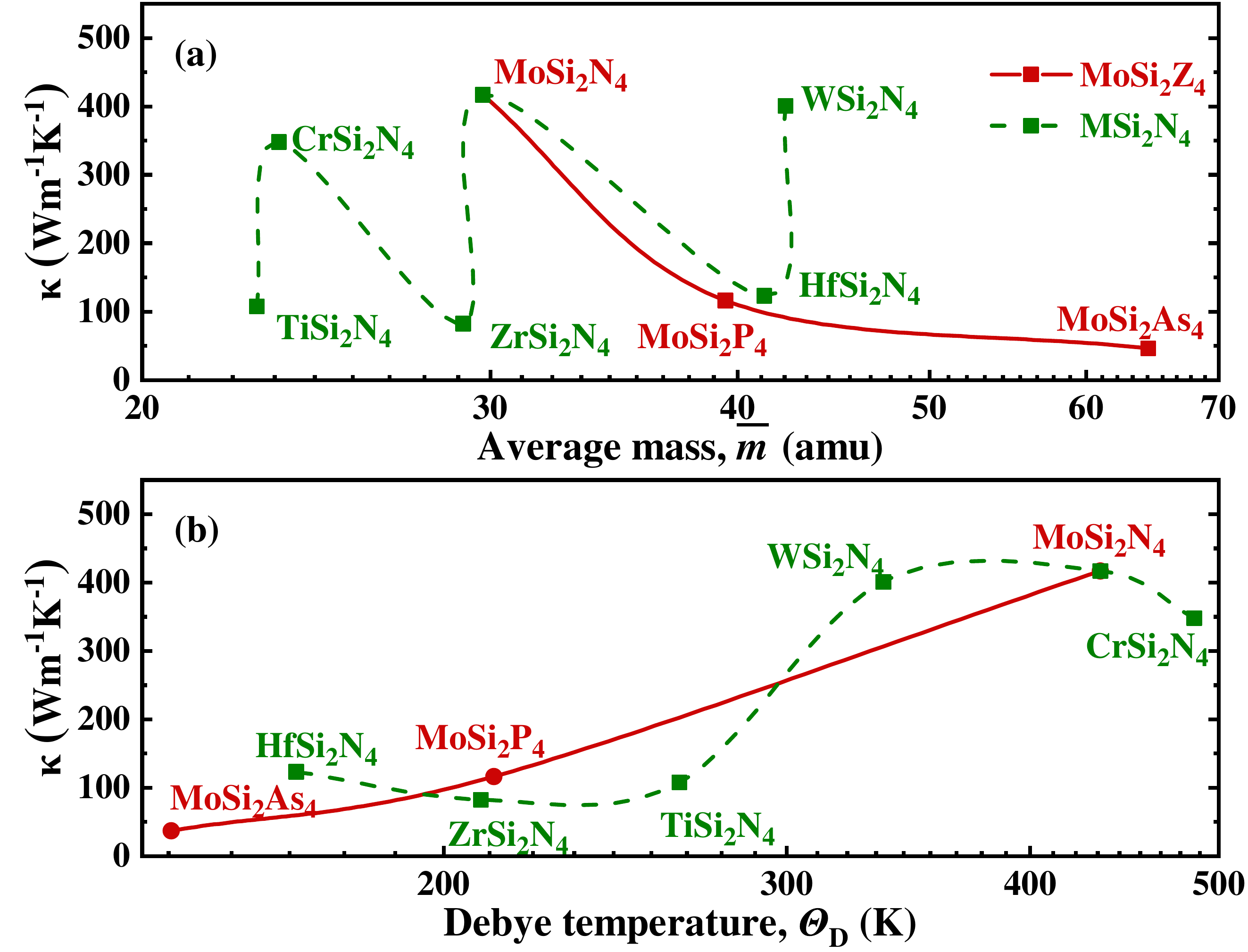}
\caption{The room-temperature thermal conductivity of MSi$_2$Z$_4$ as the function of (a) average mass $\overline{m}$ and (b) Debye temperature $\varTheta_\text{D}$.}
\label{fig4}
\end{figure}

Firstly, the large reduction of $\kappa$ with respect to Z in MoSi$_2$Z$_4$ is positively related to the increasing $\overline{m}$ and decreasing $\varTheta_\text{D}$, as shown in Fig.~\ref{fig4}. We also consider $\kappa$ in the case of only changing Z-atom mass. It is found that when N mass is substituted by P mass or As mass in MoSi$_2$N$_4$ (Fig.~S3), $\kappa$ is 70$\%$ or 46$\%$ of that in intrinsic MoSi$_2$N$_4$, respectively. This is much higher than that of intrinsic MoSi$_2$P$_4$ (28$\%$) and MoSi$_2$As$_4$ (11$\%$). The result elucidates that the atomic mass is not the only dominant factor.

\begin{figure}[!t]
\centering
\includegraphics[width=11cm]{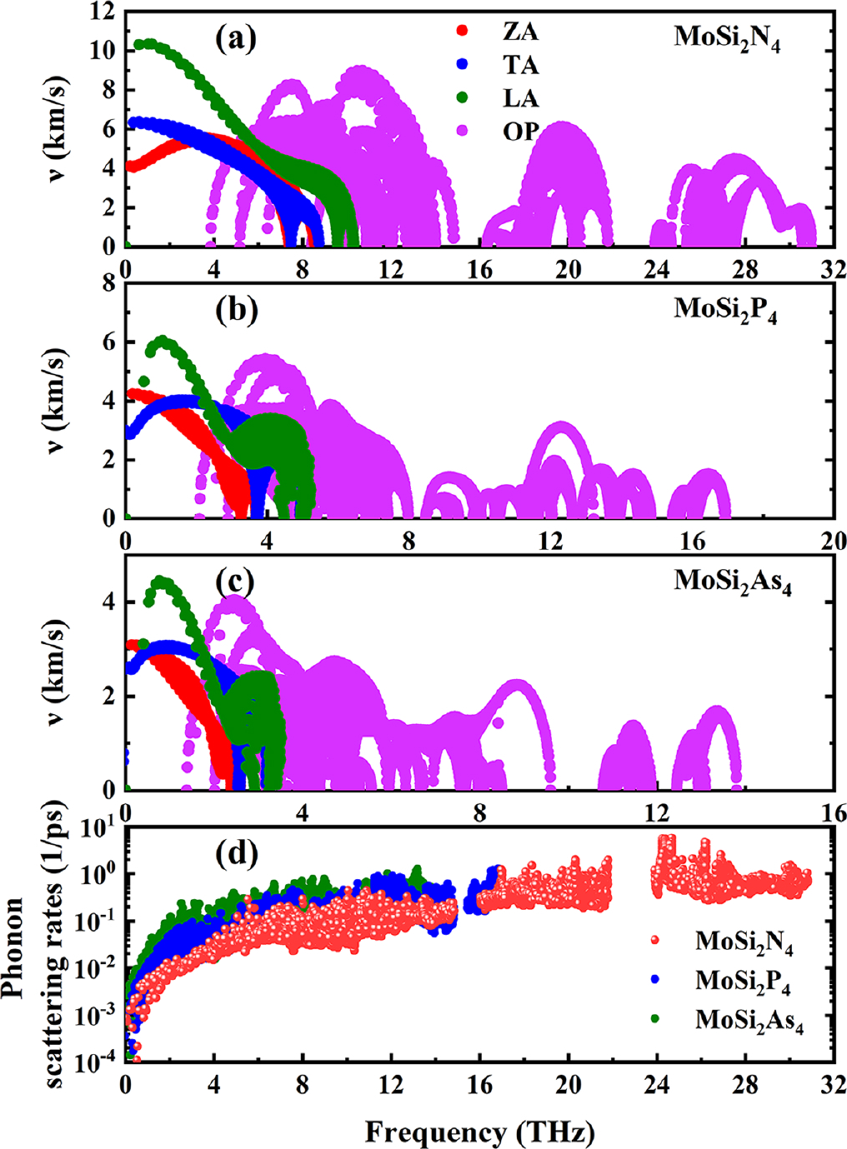}
\caption{Phonon group velocity of (a) MoSi$_2$N$_4$, (b) MoSi$_2$P$_4$ and (c) MoSi$_2$As$_4$, and (d) their comparison of phonon scattering rates (1/$\tau$).}
\label{fig5}
\end{figure}

Secondly, in phonon-dominated thermal transport, according to the Fermi's Golden rule~\cite{2009-lindsay-carbonNanotubes-PRB-FermiGR, 2019-metalKappa-Baohua-PRB}, phonon scattering processes characterized by phonon group velocity ($v$) and phonon scattering rates (1/$\tau$) could influence $\kappa$. Small $v$ and large 1/$\tau$ mean slow and short-lifetime phonons and thus low $\kappa$. Consequently, as shown in Fig.~\ref{fig5}, the decrease in $v$ and increase in 1/$\tau$ are demonstrably large for Z varying from N, P, to As in MoSi$_2$Z$_4$, thus leading to a large reduction in $\kappa$.

\begin{figure}[!t]
\centering
\includegraphics[width=11cm]{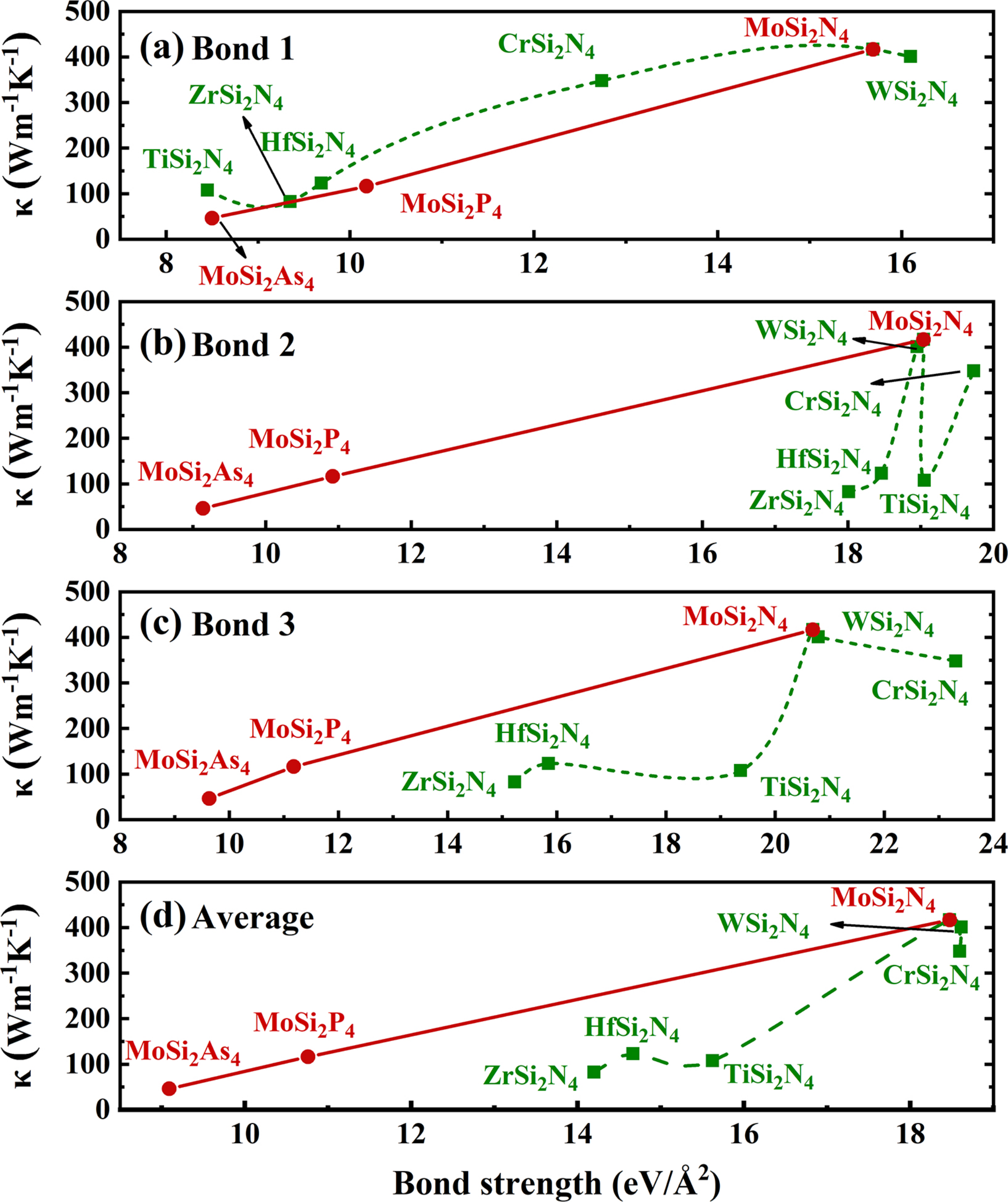}
\caption{The room-temperature thermal conductivity of MSi$_2$Z$_4$ as the function of bond strength: (a) Bond 1, (b) Bond 2, (c) Bond 3 and (d) average bond.}
\label{fig6}
\end{figure}

Thirdly, we examine the role of bond strength. From the phonon dispersion and PhDOS of the intrinsic MoSi$_2$As$_4$ and MoSi$_2$N$_4$ with As mass (Fig.~S4), we find that the maximum phonon frequency of the former is much lower than that of the latter, and in PhDOS an obvious red shift of Si/As and Mo/Si hybrid peaks occurs in the former when compared to the latter. This indicates much weak bond between As and adjacent neighbors, which can also be verified by the calculated interatomic force constants (IFCs) in Fig.~\ref{fig6}. Thus, the weaker Mo/Si-As bond strength substantially softens the phonon branches, and then results in stronger phonon-phonon scattering and thus quite smaller $\kappa$.

Fourthly, the cumulative $\kappa$ as a function of mean free path (MFP) (Fig.~S5) shows that 80$\%$ of $\kappa$ at 300~K involves the phonons with a MFP of 63.01\,$\mu$m for MoSi$_2$N$_4$, 4.31\,$\mu$m for MoSi$_2$P$_4$, and 1.35\,$\mu$m for MoSi$_2$As$_4$. Therefore, MoSi$_2$As$_4$ possesses a MFP more than one order of magnitude smaller than that in MoSi$_2$N$_4$, also explaining its quite low $\kappa$.

In short, despite of the large reduction of $\kappa$ of MoSi$_2$Z$_4$ when Z changes from N to P to As, it intrinsically obeys Slack's classic rule. Heavy atoms (e.g. As and P here) and weak bonds (e.g. Mo-As, Si-As here) could favor high phonon scattering rates, low Debye temperature, low phonon group velocity, and high anharmonicity, thus resulting in $\kappa$ of MoSi$_2$As$_4$ almost one order of magnitude smaller than that of MoSi$_2$N$_4$.

\subsection{2.4. Anomalous variation of $\kappa$ in MSi$_2$N$_4$} \label{section4}
In contrast to the variation of $\kappa$ with Z in MoSi$_2$Z$_4$ which follows Slack's rule, its variation with respect to M in MSi$_2$N$_4$ deviates from the rule of thumb proposed by Slack.

Firstly, as shown in Fig.~\ref{fig4}, Table~\ref{table1}, and Fig.~S6, $\kappa$ of MSi$_2$N$_4$ does not decrease with the increasing $\overline{m}$ (determined by M mass) and the decreasing $\varTheta_\text{D}$ and $v$, while an irregular oscillation occurs and no conspicuous relations exist.
Secondly, $\kappa$ of MSi$_2$N$_4$ remains at the same level regardless of giant difference in M mass, $\varTheta_\text{D}$ and $v$ when M is in the same group. For example, $\overline{m}$ and $\varTheta_\text{D}$ in WSi$_2$N$_4$ is 1.5 times as those in CrSi$_2$N$_4$, but $\kappa$ only shows 13$\%$ difference. This indicates that $\kappa$ of MSi$_2$N$_4$ with M in the same group is insensitive to $\overline{m}$, $\varTheta_\text{D}$ and $v$.
Thirdly, $\kappa$ of MSi$_2$N$_4$ with heavier group-\uppercase\expandafter{\romannumeral6}B M is around 400~Wm$^{-1}$K$^{-1}$, which is around three to four times over that with lighter group-\uppercase\expandafter{\romannumeral4}B M. In other words, heavy M atoms could even result in much higher $\kappa$, contradictory to Slack's rule that light atoms favour high $\kappa$.
Fourthly, in MSi$_2$N$_4$ the conventional guideline that $\varTheta_\text{D}$ decreases with increasing $\overline{m}$ holds only for M in the same group, while is invalid for M in different groups. For instance, $\varTheta_\text{D}$ of MoSi$_2$N$_4$ is much higher than that of ZrSi$_2$N$_4$ though their $\overline{m}$ is very close. This could be attributed to that ZA and TA branches of MSi$_2$N$_4$ with group-\uppercase\expandafter{\romannumeral4}B M is much flatter than that with group-\uppercase\expandafter{\romannumeral6}B M (Fig.~\ref{fig2}).
Fifthly, $\kappa$ of MSi$_2$N$_4$ does not strictly increase with the interatomic bond strength (Fig.~\ref{fig6}), deviating from Slack's rule that stronger bonds facilitate higher $\kappa$. Actually, in MSi$_2$N$_4$ here $\kappa$ increases with the average bond strength (Fig.~\ref{fig6}\,(d)) only when M belongs to different groups. The stronger bonds induce much higher $\kappa$ of MSi$_2$N$_4$ with group-\uppercase\expandafter{\romannumeral6}B M than that with group-\uppercase\expandafter{\romannumeral4}B M. However, when M is in the same group, the effect of bond strength is negligible.

In brief, $\kappa$ varying with M in MSi$_2$N$_4$ is anomalous and cannot be simply described by Slack's classic rule. In terms of M in group-\uppercase\expandafter{\romannumeral4}B and -\uppercase\expandafter{\romannumeral6}B as a whole, $\kappa$ of MSi$_2$N$_4$ neither increases with the decreasing $\overline{m}$, nor increases with the increasing $\varTheta_\text{D}$, $v$ and bond strength. It depends strongly on the group of M and remains at the same level (around 100 and 400~Wm$^{-1}$K$^{-1}$ for group-\uppercase\expandafter{\romannumeral4}B and group-\uppercase\expandafter{\romannumeral6}B M atoms, respectively) irrespective of M mass and bond strength for M in the same group.

\begin{figure}[!t]
\centering
\includegraphics[width=9.5cm]{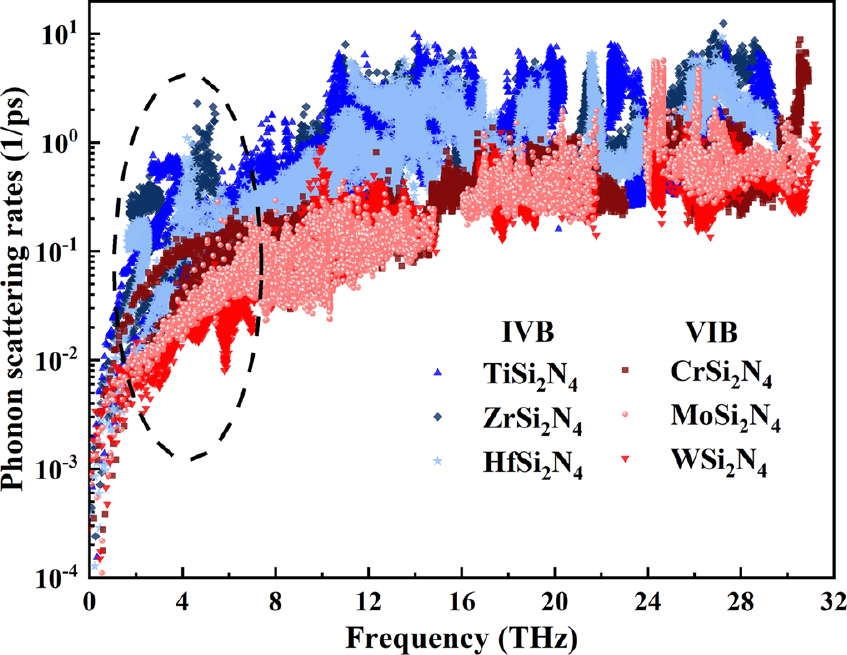}
\caption{Total phonon scattering rates of group-\uppercase\expandafter{\romannumeral4}B and -\uppercase\expandafter{\romannumeral6}B MSi$_2$N$_4$.}
\label{fig7}
\end{figure}

To this end, we trace the above abnormality to fundamental vibrational properties, which are inaccessible from conventional guideline in searching for high $\kappa$ materials. 
Figure~\ref{fig7} gives the frequency-dependent total phonon scattering rates (1/$\tau$) for MSi$_2$N$_4$. It is obvious that 1/$\tau$ in MSi$_2$N$_4$ with group-\uppercase\expandafter{\romannumeral6}B M is much smaller than that with group-\uppercase\expandafter{\romannumeral4}B M in the whole frequency range, resulting in weaker phonon scattering and thus higher $\kappa$ of the former. When M is in the same group, 1/$\tau$ is similar and thus $\kappa$ remains at the same level. There is anomalous phonon scattering as shown in the circle of Fig.~\ref{fig7}. In the frequency range of 1.5--6~THz, 1/$\tau$ of MSi$_2$N$_4$ with M in the same group abnormally decreases with the increasing M mass. This is understandable that in Fig.~S7, the phonon phase space (P3) and absolute Gr{\"u}neisen parameter ($|\gamma|$) approximately decrease with the increasing M mass, which indicates the reduced phonon scattering channels, weak scattering strengths, and thus higher $\kappa$.

The contribution of each phonon branch to the total $\kappa$ is shown in Table~\ref{table2}. One can find that the acoustic branches dominate $\kappa$ of MSi$_2$N$_4$, whose contribution is 74--84$\%$. Besides, the lowest five optical branches contribute 14--25$\%$. LA branch contributes more in the case of group-\uppercase\expandafter{\romannumeral6}B M, while TA branch dominates in the case of group-\uppercase\expandafter{\romannumeral4}B M.

\begin{table}[htbp]
\centering
\caption{Calculated $\kappa$ contributions from different phonon branches. OP~(low) indicates the first six low-frequency optical branches.}
\renewcommand\arraystretch{1.5} 
\setlength{\tabcolsep}{2mm}
\begin{tabular}{cccccc}
\toprule
 &\multicolumn{5}{c}{Thermal Contribution ($\%$)} \\
 &ZA &TA &LA &OP~(total) &OP~(low) \\  
\midrule
CrSi$_2$N$_4$ &8.7 &37.7 &35.8 &17.8 &16.8 \\
MoSi$_2$N$_4$ &21.7 &20.6 &34.5 &23.2 &22.2 \\
WSi$_2$N$_4$ &18.6 &22.1 &33.4 &25.9 &24.9 \\
TiSi$_2$N$_4$ &8.2 &57.3 &15.3 &19.2 &16.1 \\
ZrSi$_2$N$_4$ &14.3 &45.5 &23.9 &16.3 &13.8 \\
HfSi$_2$N$_4$ &11.7 &46.0 &25.8 &16.5 &14.2 \\
\bottomrule
\end{tabular}
\label{table2}
\end{table}

\begin{figure*}[!t]
\centering
\includegraphics[width=15cm]{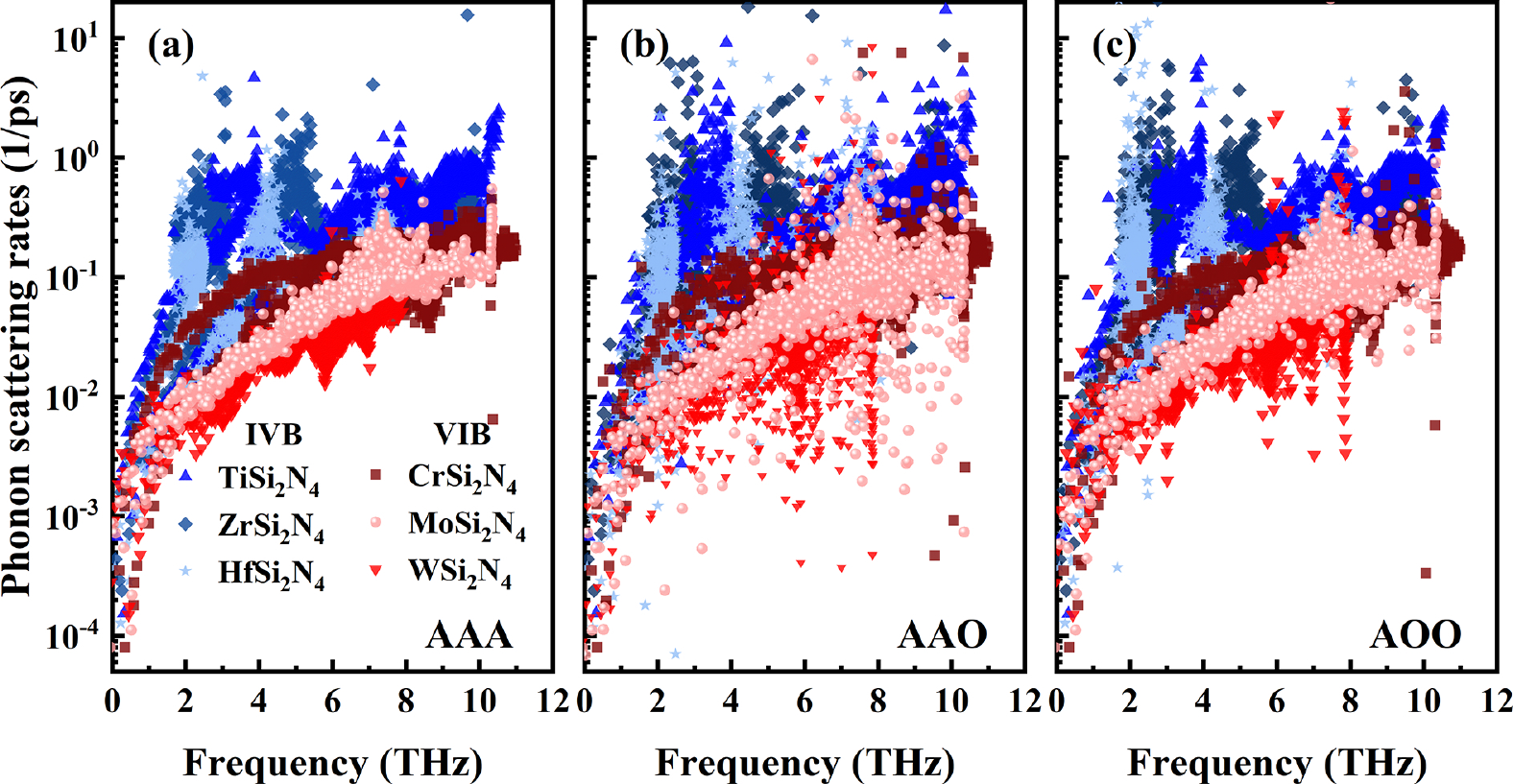}
\caption{Phonon scattering rates of MSi$_2$N$_4$ for particular interacting channels: (a) AAA, (b) AAO and (c) AOO.}
\label{fig8}
\end{figure*}

For more detailed evaluation, we examine the three-phonon scattering channels involving three acoustic phonons (AAA), two acoustic phonons combining with one optical phonon (AAO), and one acoustic phonon combining with two optical phonons (AOO), as shown in Fig.~\ref{fig8}. For each scattering channel, the scattering process is found much weaker in group-\uppercase\expandafter{\romannumeral6}B M than group-\uppercase\expandafter{\romannumeral4}B M. This is intrinsically attributed to the interplay of certain basic properties such as bunching/flatness of acoustic/optical branches, frequency gap between acoustic and optical phonons (a-o gap), etc~\cite{2012-GaN-lindsay-prl, 2013-BAs-lindsay-prl}. 
It is apparent in Fig.~\ref{fig2} that the acoustic branches in MSi$_2$N$_4$ with group-\uppercase\expandafter{\romannumeral6}B M are more bunched and less flattened than those in MSi$_2$N$_4$ with group-\uppercase\expandafter{\romannumeral4}B M. In addition, the lowest six optical branches close to acoustic branches are also more bunched in MSi$_2$N$_4$ with group-\uppercase\expandafter{\romannumeral6}B M. These give weaker AAA, AAO, and AOO scattering and thus high $\kappa$ of MSi$_2$N$_4$ with group-\uppercase\expandafter{\romannumeral6}B M. In MSi$_2$N$_4$ with M in the same group, the slight difference in the combination of these phonon features contributes to the subtle difference in $\kappa$.

The phonon scattering rates for different scattering channels also exhibit abnormality with respect to $\overline{m}$, as shown in Fig.~S8. For instance, $\overline{m}$ of MoSi$_2$N$_4$ is close to that of ZrSi$_2$N$_4$, but the rates for all the AAA, AAO, and AOO scattering in MoSi$_2$N$_4$ is one order of magnitude lower than that in ZrSi$_2$N$_4$, leading to $\kappa$ of ZrSi$_2$N$_4$ is only 20$\%$ of that of MoSi$_2$N$_4$. Though $\overline{m}$ of WSi$_2$N$_4$ is two times of that of TiSi$_2$N$_4$, $\kappa$ of the former are nearly four times of that of the latter due to the much lower rates for AAA, AAO and AOO scattering in WSi$_2$N$_4$.

\section{3. CONCLUSIONS}
In summary, we apply density functional theory and phonon Boltzmann transport equation to calculate the $\kappa$ of eight types of 2D MSi$_2$Z$_4$ semiconductors (M\,=\,Mo,\,Cr,\,W,\,Ti,\,Zr,\,Hf and Z\,=\,N,\,P,\,As) and deliver the fundamental understanding of $\kappa$. The main conclusions are summarized in the following.

1) 2D semiconductors MoSi$_2$N$_4$ and WSi$_2$N$_4$ are found to possess a high $\kappa$ over 400~Wm$^{-1}$K$^{-1}$, comparable to that of the good thermal conductor copper.

2) In MoSi$_2$Z$_4$, $\kappa$ significantly decreases by one order of magnitude from Z\,=\,N to Z\,=\,As. The variation of $\kappa$ with Z complies with Slack's classic rule. Heavy atoms (e.g. As) and weak bonds (e.g. Mo-As, Si-As) could favor high phonon scattering rates, low Debye temperature, low phonon group velocity, and high anharmonicity and thus low $\kappa$.

3) When M is in the same group, $\kappa$ of MSi$_2$N$_4$ varying with M is anomalous and deviates from Slack's classic rule. It is insensitive to $\overline{m}$, $\varTheta_\text{D}$, $v$, and bond strength owing to the similar phonon scattering properties.

4) When M is in different group, $\kappa$ of MSi$_2$N$_4$ with group-\uppercase\expandafter{\romannumeral6}B M is three to four times large as that with group-\uppercase\expandafter{\romannumeral4}B M in spite of M mass, due to that the phonon scattering rates of former are nearly one order of magnitude lower than that of the latter. This could be intrinsically ascribed to the much higher bond strength of M-N bonds and exterior Si-N bonds in MSi$_2$N$_4$ with group-\uppercase\expandafter{\romannumeral6}B M.

5) Although $\kappa$ variation in MSi$_2$N$_4$ with respect to M as a whole is anomalous and cannot be well described by Slack's rule of thumb, it could be traced to an interplay of certain basic vibrational properties that lie outside of the conventional guidelines. The acoustic branches are more bunched and less flattened, and their nearby six optical branches are more bunched in MSi$_2$N$_4$ with group-\uppercase\expandafter{\romannumeral6}B M than in MSi$_2$N$_4$ with group-\uppercase\expandafter{\romannumeral4}B M, giving weaker AAA, AAO, and AOO scattering and thus higher $\kappa$ of the former. In MSi$_2$N$_4$ with M in the same group, the slight difference in these phonon features, i.e. bunching strength and flatness of acoustic branches and their nearby optical branches, contributes to the subtle difference in $\kappa$.

In this work, it has been identified that 2D semiconductor MSi$_2$N$_4$ with group-\uppercase\expandafter{\romannumeral6}B M possesses high $\kappa$ which could open new opportunities for heat dissipation and thermal management in semiconductor device. The anomalous variation of $\kappa$ in MSi$_2$Z$_4$  has been clarified as well.
Since MSi$_2$Z$_4$ is the multi-layered 2D structure compared with other materials (e.g. graphene, silicene, BP, h-BN, TMDs etc.), the work has also given important insights into analyzing the physics of $\kappa$ in layered 2D systems. 

\section{4. METHODOLOGY SECTION}
The first-principles calculations are performed based on density functional theory (DFT)~\cite{1965Self}, as implemented in the Vienna \textit{ab initio} simulation package (VASP)~\cite{Hafner2007-vasp,Hafner2008-vasp}. The generalized gradient approximation (GGA) functional within Perdew-Burke-Ernzerhof (PBE) form is applied for the exchange-correlation functional~\cite{1996Efficient}, and the projector-augmented-wavepotential (PAW) is used to describe the electron-ion interactions~\cite{1994PAW}. The convergence criteria for energies and forces are $10^{-7}$~eV and $10^{-4}$~eV/\AA, respectively. A cutoff energy of 500 eV and 15\,$\times$\,15$\times$\,1 Monkhorst-Pack $k$-mesh in the first Brilouin zone are adopted. A vacuum layer of 20~\AA~is used to eliminate the interaction between layers.

Electron localization function (ELF) of MSi$_2$Z$_4$ is considered, as shown in Fig.~S9. 
The maximum ELF value between the two atoms of each bond is above 0.85, indicating interstitial strong localization of electrons and thus a covalent-like nature of these bonds. Considering the polarization effect of group-\uppercase\expandafter{\romannumeral4}B and -\uppercase\expandafter{\romannumeral6}B transition metal atoms on phononic nature and further thermal transport properties, we also calculate Born effective charges and dielectric constant, listed in Table S2.
Phonon dispersion and harmonic ($2^\text{nd}$) interatomic force constants (IFCs) are calculated via finite displacement method within PHONOPY package~\cite{2015phonopy}. The comparison of phonon dispersion of MoSi$_2$N$_4$ with or without polarization is shown in Fig.~S10. It is worth noting that, the high-frequency optical branches show obvious splitting near $\Gamma$ point which indicates the strong polarization phenomenon. Thus, Born effective charges and dielectric constant should be considered in calculations of phonon dispersion and thermal conductivity.
The supercell of 5\,$\times$\,5\,$\times$\,1 and $k$-mesh of 5\,$\times$\,5\,$\times$\,1 are adopted. The anharmonic ($3^\text{rd}$) IFCs are obtained by the same supercell and k-mesh. Then, $\kappa$ and phonon scattering mechanism are obtained by solving phonon Boltzmann transport equation (BTE)~\cite{2010BTE, 2014ShengBTE}. The calculating process is performed by ShengBTE package with a $q$-mesh of 81\,$\times$\,81\,$\times$\,1 and a cutoff distance of 0.5~nm. The convergence of $\kappa$ against $q$-mesh is tested from 61\,$\times$\,61\,$\times$\,1 to 101\,$\times$\,101\,$\times$\,1 as shown in Figure S11.

Lattice thermal conductivity tensor ($\kappa_{\alpha\beta}$) at different temperature ($T$) 
can be expressed as~\cite{2014ShengBTE}:

\vspace{-15pt}
\begin{equation}
\begin{split}
\kappa_{\alpha\beta}=\frac{1}{k_\text{B}T^2NV} \sum_{\lambda}(\hbar\omega_\lambda) f_\lambda^{0}(1+f_\lambda^0) \nu^\alpha_\lambda F^\beta_\lambda
\end{split}
\end{equation}

\noindent
where $k_\text{B}$, $\hbar$ and $f_{\lambda}^{0}$ are Boltzmann constant, reduced Planck constant and equilibrium Bose-Einstein distribution function, respectively. $V$ is the volume of unit cell. $\lambda$ denotes a phonon branch characterized by the wave vector $q$ points. $N$ is the number of $q$ points in the Brillouin zone. $\omega_{\lambda}$ and $\nu_{\lambda}$ are the angular frequency and phonon group velocity. $F_{\lambda}^{\beta}$ is defined as~\cite{2009-Diamond-P3}:

\vspace{-10pt}
\begin{equation}
F_\beta^\lambda=\tau_\lambda^\text{0}(\nu_\lambda^\beta+\Delta_\lambda^\beta)
\end{equation}

\noindent
in which $\tau_{\lambda}^{0}$ is phonon relaxation time from single-mode relaxation time approximation and $\Delta_{\lambda}^{\beta}$ is the correction term from iterative solution of BTE to modify the error.

In insulating and semiconductor materials, heat is mainly transported by lattice vibration~\cite{2018-MD-BTE-NEGF-ACSomega}. Here the thermal transport properties of the 2D semiconductor MSi$_2$Z$_4$ system are considered, and three-phonon scattering mechanism which contains phonon-phonon (pp) diffusion and isotope effect (iso) is discussed. The phonon relaxation time can be expressed as:

\vspace{-10pt}
\begin{equation}
\frac{\text{1}}{\tau_{\lambda}^{\text{0}}}=\frac{\text{1}}{\tau_{\lambda}^{\text{pp}}}+\frac{\text{1}}{\tau_{\lambda}^{\text{iso}}}
\end{equation}

Debye temperature is a single parameter to describe the temperature-dependent heat capacity in conventional Debye model~\cite{1997-Ge-debyeCallaway-PRB}. It depends on the dispersion law of acoustic phonons, and the average Debye temperature ($\varTheta_\text{D}$) is expressed as~\cite{2018GroupIVSe, 2016-debye-yang}:

\vspace{-10pt}
\begin{equation}
\frac{\text{1}}{\varTheta_{\text{D}}^{\text{3}}}=\frac{\text{1}}{\text{3}}(\frac{\text{1}}{\varTheta_\text{ZA}^\text{3}}+\frac{\text{1}}{\varTheta_\text{TA}^\text{3}}+\frac{\text{1}}{\varTheta_\text{TA}^\text{3}})
\end{equation}

\noindent
and $\varTheta_{i}$ ($i$\,=\,ZA,\,TA,\,LA) is defined as

\vspace{-10pt}
\begin{equation}
\varTheta_\text{$i$}=\frac{\hbar\omega_\text{$i$}}{\text{$k$}_\text{B}}
\end{equation}

\noindent
where $\omega_\text{$i$}$ is the maximum frequency of $i^\text{th}$ acoustic branch.

\section*{ACKNOWLEDGMENT}
The authors acknowledge the support from 15$^\text{th}$ Thousand Youth Talents Program of China, the National Natural Science Foundation of China (NSFC 11902150), the Research Fund of State Key Laboratory of Mechanics and Control of Mechanical Structures (MCMS-I-0419G01, MCMS-I-0421K01), the Fundamental Research Funds for the Central Universities, the open project of National Laboratory of Solid State Microstructures (M34055) and a project Funded by the Priority Academic Program Development of Jiangsu Higher Education Institutions.

\section*{AUTHOR INFORMATION}
\textbf{Corresponding Author}\\
\textbf{Min Yi} -- Institute for Frontier Science, Institute of Nanoscience, State Key Lab of Mechanics and Control of Mechanical Structures \& Key Lab for Intelligent Nano Materials and Devices of Ministry of Education \& College of Aerospace Engineering, Nanjing University of Aeronautics and Astronautics (NUAA), Nanjing 210016, China; National Laboratory of Solid State Microstructures, Nanjing University, Nanjing 210093, China;\\
Email: yimin@nuaa.edu.cn\\
\textbf{Wanlin Guo} -- Institute for Frontier Science, Institute of Nanoscience, State Key Lab of Mechanics and Control of Mechanical Structures \& Key Lab for Intelligent Nano Materials and Devices of Ministry of Education \& College of Aerospace Engineering, Nanjing University of Aeronautics and Astronautics (NUAA), Nanjing 210016, China;\\
Email: wlguo@nuaa.edu.cn\\
\textbf{Author}\\
\textbf{Yan Yin} -- Institute for Frontier Science, Institute of Nanoscience, State Key Lab of Mechanics and Control of Mechanical Structures \& Key Lab for Intelligent Nano Materials and Devices of Ministry of Education \& College of Aerospace Engineering, Nanjing University of Aeronautics and Astronautics (NUAA), Nanjing 210016, China;\\
Email: yanyin0330@nuaa.edu.cn

\section*{SUPPORTING INFORMATION}
Electronic band structures and phonon density of states of MSi$_2$Z$_4$ (Figure S1 and S2); the effect of mass and bond strength induced by Z atom on the thermal conductivity ($\kappa$) of MoSi$_2$Z$_4$ (Figure S3 and S4); the normalized cumulative thermal conductivity versus mean free path of MSi$_2$Z$_4$ (Figure S5); phonon group velocity, room-temperature $\kappa$, phase space and absoluted Gr{\"u}neisen parameter versus frequency of MSi$_2$N$_4$(Figure S6 and S7); phonon scattering rates for particular interacting channels (Figure S8); electron localization function (ELF) distribution of MSi$_2$Z$_4$ in the (110) plane (Figure S9); phonon dispersion spectrum of MoSi$_2$N$_4$ with or without polarization effect (Figure S10); convergence test of grid density of $\kappa$ (Figure S11); band gaps of 2D MSi$_2$Z$_4$ by GGA and HSE (Table S1); born effective charges and dielectric constants of 2D MSi$_2$Z$_4$ (Table S2)

\bibliography{References}

\providecommand{\latin}[1]{#1}
\makeatletter
\providecommand{\doi}
  {\begingroup\let\do\@makeother\dospecials
  \catcode`\{=1 \catcode`\}=2 \doi@aux}
\providecommand{\doi@aux}[1]{\endgroup\texttt{#1}}
\makeatother
\providecommand*\mcitethebibliography{\thebibliography}
\csname @ifundefined\endcsname{endmcitethebibliography}
  {\let\endmcitethebibliography\endthebibliography}{}
\begin{mcitethebibliography}{55}
\providecommand*\natexlab[1]{#1}
\providecommand*\mciteSetBstSublistMode[1]{}
\providecommand*\mciteSetBstMaxWidthForm[2]{}
\providecommand*\mciteBstWouldAddEndPuncttrue
  {\def\EndOfBibitem{\unskip.}}
\providecommand*\mciteBstWouldAddEndPunctfalse
  {\let\EndOfBibitem\relax}
\providecommand*\mciteSetBstMidEndSepPunct[3]{}
\providecommand*\mciteSetBstSublistLabelBeginEnd[3]{}
\providecommand*\EndOfBibitem{}
\mciteSetBstSublistMode{f}
\mciteSetBstMaxWidthForm{subitem}{(\alph{mcitesubitemcount})}
\mciteSetBstSublistLabelBeginEnd
  {\mcitemaxwidthsubitemform\space}
  {\relax}
  {\relax}

\bibitem[Koenig \latin{et~al.}(2014)Koenig, Doganov, Schmidt, Castro~Neto, and
  {\"O}zyilmaz]{2014-BlackP-electricFieldEffect-apl}
Koenig,~S.~P.; Doganov,~R.~A.; Schmidt,~H.; Castro~Neto,~A.~H.;
  {\"O}zyilmaz,~B. Electric Field Effect in Ultrathin Black Phosphorus.
  \emph{Appl. Phys. Lett.} \textbf{2014}, \emph{104}, 103106\relax
\mciteBstWouldAddEndPuncttrue
\mciteSetBstMidEndSepPunct{\mcitedefaultmidpunct}
{\mcitedefaultendpunct}{\mcitedefaultseppunct}\relax
\EndOfBibitem
\bibitem[Gu and Yang(2014)Gu, and Yang]{2014-TMDs-APL-guxiaokun}
Gu,~X.; Yang,~R. Phonon Transport in Single-layer Transition Metal
  Dichalcogenides: A First-principles Study. \emph{Appl. Phys. Lett.}
  \textbf{2014}, \emph{105}, 131903\relax
\mciteBstWouldAddEndPuncttrue
\mciteSetBstMidEndSepPunct{\mcitedefaultmidpunct}
{\mcitedefaultendpunct}{\mcitedefaultseppunct}\relax
\EndOfBibitem
\bibitem[Qin \latin{et~al.}(2016)Qin, Qin, Fang, Zhang, Yue, Yan, Hu, and
  Su]{2016GroupIV-VI-A}
Qin,~G.; Qin,~Z.; Fang,~W.; Zhang,~L.; Yue,~S.; Yan,~Q.; Hu,~M.; Su,~G. Diverse
  Anisotropy of Phonon Transport in Two-dimensional Group
  \uppercase\expandafter{\romannumeral4} -
  \uppercase\expandafter{\romannumeral6} Compounds: A Comparative Study.
  \emph{Nanoscale} \textbf{2016}, \emph{8}, 11306--11319\relax
\mciteBstWouldAddEndPuncttrue
\mciteSetBstMidEndSepPunct{\mcitedefaultmidpunct}
{\mcitedefaultendpunct}{\mcitedefaultseppunct}\relax
\EndOfBibitem
\bibitem[Liu \latin{et~al.}(2018)Liu, Bo, Xu, Yin, Zhang, Wang, Eriksson, and
  Wang]{2018GroupIVSe}
Liu,~P.; Bo,~T.; Xu,~J.; Yin,~W.; Zhang,~J.; Wang,~F.; Eriksson,~O.;
  Wang,~B.~T. First-principles Calculations of the Ultralow Thermal
  Conductivity in Two-dimensional Group-\uppercase\expandafter{\romannumeral4}
  Selenides. \emph{Phys. Rev. B} \textbf{2018}, \emph{98}, 235426\relax
\mciteBstWouldAddEndPuncttrue
\mciteSetBstMidEndSepPunct{\mcitedefaultmidpunct}
{\mcitedefaultendpunct}{\mcitedefaultseppunct}\relax
\EndOfBibitem
\bibitem[Taube \latin{et~al.}(2015)Taube, Judek, {\L}api\'nska, and
  Zdrojek]{2015-MoS2-Au-InterfaceResistance-ACSAMI}
Taube,~A.; Judek,~J.; {\L}api\'nska,~A.; Zdrojek,~M. Temperature-dependent
  Thermal Properties of Supported {MoS$_2$} Monolayers. \emph{ACS Appl. Mater.
  Inter.} \textbf{2015}, \emph{7}, 5061--5065\relax
\mciteBstWouldAddEndPuncttrue
\mciteSetBstMidEndSepPunct{\mcitedefaultmidpunct}
{\mcitedefaultendpunct}{\mcitedefaultseppunct}\relax
\EndOfBibitem
\bibitem[Zhang \latin{et~al.}(2015)Zhang, Sun, Li, Lee, Cui, Chenet, You,
  Heinz, and Hone]{2015-MoS2-MoSe2-Interfacial-ACSAMI}
Zhang,~X.; Sun,~D.; Li,~Y.; Lee,~G.; Cui,~X.; Chenet,~D.; You,~Y.;
  Heinz,~T.~F.; Hone,~J.~C. Measurement of Lateral and Interfacial Thermal
  Conductivity of Single- and Bilayer {MoS$_2$} and {MoSe$_2$} Using Refined
  Optothermal Raman Technique. \emph{ACS Appl. Mater. Inter.} \textbf{2015},
  \emph{7}, 25923--25929\relax
\mciteBstWouldAddEndPuncttrue
\mciteSetBstMidEndSepPunct{\mcitedefaultmidpunct}
{\mcitedefaultendpunct}{\mcitedefaultseppunct}\relax
\EndOfBibitem
\bibitem[Zhao \latin{et~al.}(2014)Zhao, Lo, Zhang, Sun, Tan, Uher, Wolverton,
  Dravid, and Kanatzidis]{2014-SnSe-Thermoelectric}
Zhao,~L.; Lo,~S.; Zhang,~Y.; Sun,~H.; Tan,~G.; Uher,~C.; Wolverton,~C.;
  Dravid,~V.~P.; Kanatzidis,~M.~G. Ultralow Thermal Conductivity and High
  Thermoelectric Figure of Merit in {SnSe} Crystals. \emph{Nature}
  \textbf{2014}, \emph{508}, 373--377\relax
\mciteBstWouldAddEndPuncttrue
\mciteSetBstMidEndSepPunct{\mcitedefaultmidpunct}
{\mcitedefaultendpunct}{\mcitedefaultseppunct}\relax
\EndOfBibitem
\bibitem[Saito \latin{et~al.}(2016)Saito, Iizuka, Koretsune, Arita, Shimizu,
  and Iwasa]{2016-BlackP-Thermoelectric}
Saito,~Y.; Iizuka,~T.; Koretsune,~T.; Arita,~R.; Shimizu,~S.; Iwasa,~Y.
  Gate-Tuned Thermoelectric Power in Black Phosphorus. \emph{Nano Lett.}
  \textbf{2016}, \emph{16}, 4819--4824\relax
\mciteBstWouldAddEndPuncttrue
\mciteSetBstMidEndSepPunct{\mcitedefaultmidpunct}
{\mcitedefaultendpunct}{\mcitedefaultseppunct}\relax
\EndOfBibitem
\bibitem[Li \latin{et~al.}(2014)Li, Yu, Ye, Ge, Ou, Wu, Feng, Chen, and
  Zhang]{2014-BlackP-fieldEffectTransistors-NatureNanotech}
Li,~L.; Yu,~Y.; Ye,~G.~J.; Ge,~Q.; Ou,~X.; Wu,~H.; Feng,~D.; Chen,~X.~H.;
  Zhang,~Y. Black Phosphorus Field-effect Transistors. \emph{Nat. Nanotechnol.}
  \textbf{2014}, \emph{9}, 372--377\relax
\mciteBstWouldAddEndPuncttrue
\mciteSetBstMidEndSepPunct{\mcitedefaultmidpunct}
{\mcitedefaultendpunct}{\mcitedefaultseppunct}\relax
\EndOfBibitem
\bibitem[Zhang \latin{et~al.}(2017)Zhang, Xie, Ouyang, and
  Chen]{2017-TMD-InternationalJ.HeatMassTransfer}
Zhang,~Z.; Xie,~Y.; Ouyang,~Y.; Chen,~Y. A Systematic Investigation of Thermal
  Conductivities of Transition Metal Dichalcogenides. \emph{Int. J. Heat Mass
  Tran.} \textbf{2017}, \emph{108}, 417--422\relax
\mciteBstWouldAddEndPuncttrue
\mciteSetBstMidEndSepPunct{\mcitedefaultmidpunct}
{\mcitedefaultendpunct}{\mcitedefaultseppunct}\relax
\EndOfBibitem
\bibitem[Balandin \latin{et~al.}(2008)Balandin, Ghosh, Bao, Calizo, and
  Lau]{2008Superior}
Balandin,~A.~A.; Ghosh,~S.; Bao,~W.; Calizo,~I.; Lau,~C. Superior Thermal
  Conductivity of Single-layer Graphene. \emph{Nano Lett.} \textbf{2008},
  \emph{8}, 902--907\relax
\mciteBstWouldAddEndPuncttrue
\mciteSetBstMidEndSepPunct{\mcitedefaultmidpunct}
{\mcitedefaultendpunct}{\mcitedefaultseppunct}\relax
\EndOfBibitem
\bibitem[Fan \latin{et~al.}(2019)Fan, Wu, Lindsay, and Hu]{2019-BN-BP-BAs-BSb}
Fan,~H.; Wu,~H.; Lindsay,~L.; Hu,~Y. Ab Initio Investigation of Single-layer
  High Thermal Conductivity Boron Compounds. \emph{Phys. Rev. B} \textbf{2019},
  \emph{100}, 085420\relax
\mciteBstWouldAddEndPuncttrue
\mciteSetBstMidEndSepPunct{\mcitedefaultmidpunct}
{\mcitedefaultendpunct}{\mcitedefaultseppunct}\relax
\EndOfBibitem
\bibitem[Hong \latin{et~al.}(2018)Hong, Zhang, and Zeng]{2018-C3N-nanoscale}
Hong,~Y.; Zhang,~J.; Zeng,~X. Monolayer and Bilayer Polyaniline {C$_3$N}:
  Two-dimensional Semiconductors With High Thermal Conductivity.
  \emph{Nanoscale} \textbf{2018}, \emph{10}, 4301--4310\relax
\mciteBstWouldAddEndPuncttrue
\mciteSetBstMidEndSepPunct{\mcitedefaultmidpunct}
{\mcitedefaultendpunct}{\mcitedefaultseppunct}\relax
\EndOfBibitem
\bibitem[Dong \latin{et~al.}(2018)Dong, Zhang, Meng, Groves, and
  Lin]{2018-C2N2-carbon}
Dong,~Y.; Zhang,~C.; Meng,~M.; Groves,~M.~M.; Lin,~J. Novel Two-dimensional
  Diamond Like Carbon Nitrides with Extraordinary Elasticity and Thermal
  Conductivity. \emph{Carbon} \textbf{2018}, \emph{138}, 319--324\relax
\mciteBstWouldAddEndPuncttrue
\mciteSetBstMidEndSepPunct{\mcitedefaultmidpunct}
{\mcitedefaultendpunct}{\mcitedefaultseppunct}\relax
\EndOfBibitem
\bibitem[Mortazavi \latin{et~al.}(2019)Mortazavi, Shahrokhi, Raeisi, Zhuang,
  Pereira, and Rabczuk]{2019-BC3-carbon}
Mortazavi,~B.; Shahrokhi,~M.; Raeisi,~M.; Zhuang,~X.; Pereira,~L. F.~C.;
  Rabczuk,~T. Outstanding Strength, Optical Characteristics and Thermal
  Conductivity of Graphene-like {BC$_3$} and {BC$_6$N} Semiconductors.
  \emph{Carbon} \textbf{2019}, \emph{149}, 733--742\relax
\mciteBstWouldAddEndPuncttrue
\mciteSetBstMidEndSepPunct{\mcitedefaultmidpunct}
{\mcitedefaultendpunct}{\mcitedefaultseppunct}\relax
\EndOfBibitem
\bibitem[Hu \latin{et~al.}(2020)Hu, Li, Yin, Li, Ding, Zhou, and
  Zhang]{2020-HYX-diamane}
Hu,~Y.; Li,~D.; Yin,~Y.; Li,~S.; Ding,~G.; Zhou,~H.; Zhang,~G. The Important
  Role of Strain on Phonon Hydrodynamics in Diamond-like Bi-layer Graphene.
  \emph{Nanotechnology} \textbf{2020}, \emph{31}, 335711\relax
\mciteBstWouldAddEndPuncttrue
\mciteSetBstMidEndSepPunct{\mcitedefaultmidpunct}
{\mcitedefaultendpunct}{\mcitedefaultseppunct}\relax
\EndOfBibitem
\bibitem[Hong \latin{et~al.}(2020)Hong, Liu, Wang, Zhou, and Ren]{2020Chemical}
Hong,~Y.; Liu,~Z.; Wang,~L.; Zhou,~T.; Ren,~W. Chemical Vapor Deposition of
  Layered Two-dimensional {MoSi$_2$N$_4$} Materials. \emph{Science}
  \textbf{2020}, \emph{369}, 670--674\relax
\mciteBstWouldAddEndPuncttrue
\mciteSetBstMidEndSepPunct{\mcitedefaultmidpunct}
{\mcitedefaultendpunct}{\mcitedefaultseppunct}\relax
\EndOfBibitem
\bibitem[Wang \latin{et~al.}(2021)Wang, Shi, Liu, Hong, Chen, Li, Gao, Ren,
  Cheng, Li, and Chen]{2021-MAZ-NatureCommunication}
Wang,~L.; Shi,~Y.; Liu,~M.; Hong,~Y.; Chen,~M.; Li,~R.; Gao,~Q.; Ren,~W.;
  Cheng,~H.; Li,~Y.; Chen,~X. Intercalated Architecture of {MA$_2$Z$_4$} Family
  Layered Van Der Waals Materials with Emerging Topological, Magnetic and
  Superconducting Properties. \emph{Nat. Commun.} \textbf{2021}, \emph{12},
  1--10\relax
\mciteBstWouldAddEndPuncttrue
\mciteSetBstMidEndSepPunct{\mcitedefaultmidpunct}
{\mcitedefaultendpunct}{\mcitedefaultseppunct}\relax
\EndOfBibitem
\bibitem[Mortazavi \latin{et~al.}(2021)Mortazavi, Javvaji, Shojaei, Rabczuk,
  V., and Zhuang]{MORTAZAVI2021-NanoEnergy}
Mortazavi,~B.; Javvaji,~B.; Shojaei,~F.; Rabczuk,~T.; V.,~S.~A.; Zhuang,~X.
  Exceptional Piezoelectricity, High Thermal Conductivity and Stiffness and
  Promising Photocatalysis in Two-dimensional {MoSi$_2$N$_4$} Family Confirmed
  by First-principles. \emph{Nano Energy} \textbf{2021}, \emph{82},
  105716\relax
\mciteBstWouldAddEndPuncttrue
\mciteSetBstMidEndSepPunct{\mcitedefaultmidpunct}
{\mcitedefaultendpunct}{\mcitedefaultseppunct}\relax
\EndOfBibitem
\bibitem[Li \latin{et~al.}(2020)Li, Wu, Feng, Guan, Feng, Yao, and
  Yang]{PhysRevB.Valley-dependent}
Li,~S.; Wu,~W.; Feng,~X.; Guan,~S.; Feng,~W.; Yao,~Y.; Yang,~S.~A.
  Valley-dependent Properties of Monolayer {MoSi$_2$N$_4$}, {WSi$_2$N$_4$}, and
  {MoSi$_2$As$_4$}. \emph{Phys. Rev. B} \textbf{2020}, \emph{102}, 235435\relax
\mciteBstWouldAddEndPuncttrue
\mciteSetBstMidEndSepPunct{\mcitedefaultmidpunct}
{\mcitedefaultendpunct}{\mcitedefaultseppunct}\relax
\EndOfBibitem
\bibitem[Yang \latin{et~al.}(2021)Yang, Song, Sun, and
  Lu]{PRB.ValleyPseudospin}
Yang,~C.; Song,~Z.; Sun,~X.; Lu,~J. Valley Pseudospin in Monolayer
  {MoSi$_2$N$_4$} and {MoSi$_2$As$_4$}. \emph{Phys. Rev. B} \textbf{2021},
  \emph{103}, 035308\relax
\mciteBstWouldAddEndPuncttrue
\mciteSetBstMidEndSepPunct{\mcitedefaultmidpunct}
{\mcitedefaultendpunct}{\mcitedefaultseppunct}\relax
\EndOfBibitem
\bibitem[Zhong \latin{et~al.}(2021)Zhong, Xiong, Lv, Yu, and
  Yuan]{PhysRevB.strain-electrical}
Zhong,~H.; Xiong,~W.; Lv,~P.; Yu,~J.; Yuan,~S. Strain-induced Semiconductor to
  Metal Transition in {MA$_2$Z$_4$} Bilayers ({M=Ti,Cr,Mo}; {A=Si}; {Z=N,P}).
  \emph{Phys. Rev. B} \textbf{2021}, \emph{103}, 085124\relax
\mciteBstWouldAddEndPuncttrue
\mciteSetBstMidEndSepPunct{\mcitedefaultmidpunct}
{\mcitedefaultendpunct}{\mcitedefaultseppunct}\relax
\EndOfBibitem
\bibitem[Cao \latin{et~al.}(2021)Cao, Zhou, Wang, Ang, and
  Ang]{apl-electricalContact-MoSi2N4}
Cao,~L.; Zhou,~G.; Wang,~Q.; Ang,~L.~K.; Ang,~Y.~S. Two-dimensional Van Der
  {Waals} Electrical Contact to Monolayer {MoSi$_2$N$_4$}. \emph{Appl. Phys.
  Lett.} \textbf{2021}, \emph{118}, 013106\relax
\mciteBstWouldAddEndPuncttrue
\mciteSetBstMidEndSepPunct{\mcitedefaultmidpunct}
{\mcitedefaultendpunct}{\mcitedefaultseppunct}\relax
\EndOfBibitem
\bibitem[Chen and Tang(2021)Chen, and
  Tang]{2021-MA2Z4-ElectronicMagneticCatalytic}
Chen,~J.; Tang,~Q. The Versatile Electronic, Magnetic and Photo-Electro
  Catalytic Activity of a New {2D} {MA$_2$Z$_4$} Family. \emph{Chem. Eur. J.}
  \textbf{2021}, \emph{27}, 1--10\relax
\mciteBstWouldAddEndPuncttrue
\mciteSetBstMidEndSepPunct{\mcitedefaultmidpunct}
{\mcitedefaultendpunct}{\mcitedefaultseppunct}\relax
\EndOfBibitem
\bibitem[Yang \latin{et~al.}(2021)Yang, Zhao, Li, Liu, Wang, Chen, Gao, and
  Zhao]{2021-MAZ-Nanoscale}
Yang,~J.; Zhao,~L.; Li,~S.; Liu,~H.; Wang,~L.; Chen,~M.; Gao,~J.; Zhao,~J.
  Accurate Electronic Properties and Non-linear Optical Response of
  Two-dimensional {MA$_2$Z$_4$}. \emph{Nanoscale} \textbf{2021}, \emph{13},
  5479--5488\relax
\mciteBstWouldAddEndPuncttrue
\mciteSetBstMidEndSepPunct{\mcitedefaultmidpunct}
{\mcitedefaultendpunct}{\mcitedefaultseppunct}\relax
\EndOfBibitem
\bibitem[Yu \latin{et~al.}(2021)Yu, Zhou, Wan, and Li]{MoSiN-WSiN-DFT-k_2021}
Yu,~J.; Zhou,~J.; Wan,~X.; Li,~Q. High Intrinsic Lattice Thermal Conductivity
  in Monolayer {MoSi$_2$N$_4$}. \emph{New J. Phys.} \textbf{2021}, \emph{23},
  033005\relax
\mciteBstWouldAddEndPuncttrue
\mciteSetBstMidEndSepPunct{\mcitedefaultmidpunct}
{\mcitedefaultendpunct}{\mcitedefaultseppunct}\relax
\EndOfBibitem
\bibitem[Slack(1962)]{1962-slack}
Slack,~G.~A. Anisotropic Thermal Conductivity of Pyrolytic Graphite.
  \emph{Phys. Rev.} \textbf{1962}, \emph{127}, 694--701\relax
\mciteBstWouldAddEndPuncttrue
\mciteSetBstMidEndSepPunct{\mcitedefaultmidpunct}
{\mcitedefaultendpunct}{\mcitedefaultseppunct}\relax
\EndOfBibitem
\bibitem[Slack(1973)]{1973-Slack}
Slack,~G.~A. Nonmetallic Crystals with High Thermal Conductivity. \emph{J.
  Phys. Chem. Solids} \textbf{1973}, \emph{34}, 321--335\relax
\mciteBstWouldAddEndPuncttrue
\mciteSetBstMidEndSepPunct{\mcitedefaultmidpunct}
{\mcitedefaultendpunct}{\mcitedefaultseppunct}\relax
\EndOfBibitem
\bibitem[Hu \latin{et~al.}(2021)Hu, Yin, Ding, Liu, Zhou, Feng, Zhang, and
  Li]{2021-hyx-biBAs-mtp}
Hu,~Y.; Yin,~Y.; Ding,~G.; Liu,~J.; Zhou,~H.; Feng,~W.; Zhang,~G.; Li,~D. High
  Thermal Conductivity in Covalently Bonded Bi-layer Honeycomb Boron Arsenide.
  \emph{Mater. Today Phys.} \textbf{2021}, \emph{17}, 100346\relax
\mciteBstWouldAddEndPuncttrue
\mciteSetBstMidEndSepPunct{\mcitedefaultmidpunct}
{\mcitedefaultendpunct}{\mcitedefaultseppunct}\relax
\EndOfBibitem
\bibitem[Li \latin{et~al.}(2020)Li, Nie, and Sun]{2020BHBFBCl}
Li,~T.; Nie,~G.; Sun,~Q. Highly Sensitive Tuning of Lattice Thermal
  Conductivity of Graphene-like Borophene by Fluorination and Chlorination.
  \emph{Nano Res.} \textbf{2020}, \emph{13}, 1171--1177\relax
\mciteBstWouldAddEndPuncttrue
\mciteSetBstMidEndSepPunct{\mcitedefaultmidpunct}
{\mcitedefaultendpunct}{\mcitedefaultseppunct}\relax
\EndOfBibitem
\bibitem[Li \latin{et~al.}(2020)Li, Liu, Zhang, Shi, Jiang, Chen, Yin, and
  Wang]{2020-GaGeTe}
Li,~J.; Liu,~P.~F.; Zhang,~C.; Shi,~X.; Jiang,~S.; Chen,~W.; Yin,~H.;
  Wang,~B.~T. {Lattice Vibrational Modes and Phonon Thermal Conductivity of
  Single-layer GaGeTe}. \emph{Journal of Materiomics} \textbf{2020}, \emph{6},
  723--728\relax
\mciteBstWouldAddEndPuncttrue
\mciteSetBstMidEndSepPunct{\mcitedefaultmidpunct}
{\mcitedefaultendpunct}{\mcitedefaultseppunct}\relax
\EndOfBibitem
\bibitem[Cai \latin{et~al.}(2014)Cai, Lan, Zhang, and Zhang]{2014MoS2}
Cai,~Y.; Lan,~J.; Zhang,~G.; Zhang,~Y. Lattice Vibrational Modes and Phonon
  Thermal Conductivity of Monolayer {MoS$_2$}. \emph{Phys. Rev. B}
  \textbf{2014}, \emph{89}, 035438\relax
\mciteBstWouldAddEndPuncttrue
\mciteSetBstMidEndSepPunct{\mcitedefaultmidpunct}
{\mcitedefaultendpunct}{\mcitedefaultseppunct}\relax
\EndOfBibitem
\bibitem[Gu and Yang(2014)Gu, and Yang]{2014-TMD-GuXiaoKun}
Gu,~X.; Yang,~R. Phonon Transport in Single-layer Transition Metal
  Dichalcogenides: A First-principles Study. \emph{Appl. Phys. Letters.}
  \textbf{2014}, \emph{105}, 131903\relax
\mciteBstWouldAddEndPuncttrue
\mciteSetBstMidEndSepPunct{\mcitedefaultmidpunct}
{\mcitedefaultendpunct}{\mcitedefaultseppunct}\relax
\EndOfBibitem
\bibitem[Torres \latin{et~al.}(2019)Torres, Alvarez, Cartoix{\`{a} X. }, and
  Rurali]{2019-TMD}
Torres,~P.; Alvarez,~F.~X.; Cartoix{\`{a} X. },; Rurali,~R. Thermal
  Conductivity and Phonon Hydrodynamics in Transition Metal Dichalcogenides
  from First-principles. \emph{2D Mater.} \textbf{2019}, \emph{6}, 035002\relax
\mciteBstWouldAddEndPuncttrue
\mciteSetBstMidEndSepPunct{\mcitedefaultmidpunct}
{\mcitedefaultendpunct}{\mcitedefaultseppunct}\relax
\EndOfBibitem
\bibitem[Sun \latin{et~al.}(2019)Sun, Shuai, and Wang]{2019-SnSe}
Sun,~Y.; Shuai,~Z.; Wang,~D. Reducing Lattice Thermal Conductivity of the
  Thermoelectric SnSe Monolayer: Role of Phonon–electron Coupling. \emph{J.
  Phys. Chem. C} \textbf{2019}, \emph{123}, 12001--12006\relax
\mciteBstWouldAddEndPuncttrue
\mciteSetBstMidEndSepPunct{\mcitedefaultmidpunct}
{\mcitedefaultendpunct}{\mcitedefaultseppunct}\relax
\EndOfBibitem
\bibitem[Ghosh \latin{et~al.}(2008)Ghosh, Calizo, Teweldebrhan, Pokatilov,
  Nika, Balandin, Bao, Miao, and Lau]{2008Graphene3000-APL}
Ghosh,~S.; Calizo,~I.; Teweldebrhan,~D.; Pokatilov,~E.~P.; Nika,~D.;
  Balandin,~A.~A.; Bao,~W.; Miao,~F.; Lau,~C.~N. Extremely High Thermal
  Conductivity of Graphene: Prospects for Thermal Management Applications in
  Nanoelectronic Circuits. \emph{Appl. Phys. Lett.} \textbf{2008}, \emph{92},
  1148\relax
\mciteBstWouldAddEndPuncttrue
\mciteSetBstMidEndSepPunct{\mcitedefaultmidpunct}
{\mcitedefaultendpunct}{\mcitedefaultseppunct}\relax
\EndOfBibitem
\bibitem[Hong \latin{et~al.}(2018)Hong, Zhang, and Zeng]{2018MonolayerC3N}
Hong,~Y.; Zhang,~J.; Zeng,~X. Monolayer and Bilayer Polyaniline {C$_3$N}:
  Two-dimensional Semiconductors with High Thermal Conductivity.
  \emph{Nanoscale} \textbf{2018}, \emph{10}, 4301--4310\relax
\mciteBstWouldAddEndPuncttrue
\mciteSetBstMidEndSepPunct{\mcitedefaultmidpunct}
{\mcitedefaultendpunct}{\mcitedefaultseppunct}\relax
\EndOfBibitem
\bibitem[Lindsay and Broido(2011)Lindsay, and Broido]{2011BN600}
Lindsay,~L.; Broido,~D.~A. Enhanced Thermal Conductivity and Isotope Effect in
  Single-layer Hexagonal Boron Nitride. \emph{Phys. Rev. B} \textbf{2011},
  \emph{84}, 155421--155421\relax
\mciteBstWouldAddEndPuncttrue
\mciteSetBstMidEndSepPunct{\mcitedefaultmidpunct}
{\mcitedefaultendpunct}{\mcitedefaultseppunct}\relax
\EndOfBibitem
\bibitem[Lindsay \latin{et~al.}(2009)Lindsay, Broido, and
  Mingo]{2009-lindsay-carbonNanotubes-PRB-FermiGR}
Lindsay,~L.; Broido,~D.~A.; Mingo,~N. Lattice Thermal Conductivity of
  Single-walled Carbon Nanotubes: Beyond the Relaxation Time Approximation and
  Phonon-phonon Scattering Selection Rules. \emph{Phys. Rev. B} \textbf{2009},
  \emph{80}, 125407\relax
\mciteBstWouldAddEndPuncttrue
\mciteSetBstMidEndSepPunct{\mcitedefaultmidpunct}
{\mcitedefaultendpunct}{\mcitedefaultseppunct}\relax
\EndOfBibitem
\bibitem[Tong \latin{et~al.}(2019)Tong, Li, Ruan, and
  Bao]{2019-metalKappa-Baohua-PRB}
Tong,~Z.; Li,~S.; Ruan,~X.; Bao,~H. Comprehensive First-principles Analysis of
  Phonon Thermal Conductivity and Electron-phonon Coupling in Different Metals.
  \emph{Phys. Rev. B} \textbf{2019}, \emph{100}, 144306\relax
\mciteBstWouldAddEndPuncttrue
\mciteSetBstMidEndSepPunct{\mcitedefaultmidpunct}
{\mcitedefaultendpunct}{\mcitedefaultseppunct}\relax
\EndOfBibitem
\bibitem[Lindsay \latin{et~al.}(2012)Lindsay, Broido, and
  Reinecke]{2012-GaN-lindsay-prl}
Lindsay,~L.; Broido,~D.~A.; Reinecke,~T.~L. Thermal Conductivity and Large
  Isotope Effect in {GaN} from First Principles. \emph{Phys. Rev. Lett.}
  \textbf{2012}, \emph{109}, 095901\relax
\mciteBstWouldAddEndPuncttrue
\mciteSetBstMidEndSepPunct{\mcitedefaultmidpunct}
{\mcitedefaultendpunct}{\mcitedefaultseppunct}\relax
\EndOfBibitem
\bibitem[Lindsay \latin{et~al.}(2013)Lindsay, Broido, and
  Reinecke]{2013-BAs-lindsay-prl}
Lindsay,~L.; Broido,~D.~A.; Reinecke,~T.~L. First-Principles Determination of
  Ultrahigh Thermal Conductivity of Boron Arsenide: A Competitor for Diamond?
  \emph{Phys. Rev. Lett.} \textbf{2013}, \emph{111}, 025901\relax
\mciteBstWouldAddEndPuncttrue
\mciteSetBstMidEndSepPunct{\mcitedefaultmidpunct}
{\mcitedefaultendpunct}{\mcitedefaultseppunct}\relax
\EndOfBibitem
\bibitem[Kohn and Sham(1965)Kohn, and Sham]{1965Self}
Kohn,~W.; Sham,~L.~J. Self-Consistent Equations Including Exchange and
  Correlation Effects. \emph{Phys. Rev.} \textbf{1965}, \emph{140}, A1133\relax
\mciteBstWouldAddEndPuncttrue
\mciteSetBstMidEndSepPunct{\mcitedefaultmidpunct}
{\mcitedefaultendpunct}{\mcitedefaultseppunct}\relax
\EndOfBibitem
\bibitem[Hafner(2007)]{Hafner2007-vasp}
Hafner,~J. {Materials Simulations using VASP-A Quantum Perspective to Materials
  Science}. \emph{Comput. Phys. Commun.} \textbf{2007}, \emph{177}, 6--13\relax
\mciteBstWouldAddEndPuncttrue
\mciteSetBstMidEndSepPunct{\mcitedefaultmidpunct}
{\mcitedefaultendpunct}{\mcitedefaultseppunct}\relax
\EndOfBibitem
\bibitem[Hafner(2008)]{Hafner2008-vasp}
Hafner,~J. {Ab-Initio Simulations of Materials Using VASP: Density-functional
  Theory and Beyond J{\"{U}}RGEN}. \emph{J. Comput. Chem.} \textbf{2008},
  \emph{29}, 2044--2078\relax
\mciteBstWouldAddEndPuncttrue
\mciteSetBstMidEndSepPunct{\mcitedefaultmidpunct}
{\mcitedefaultendpunct}{\mcitedefaultseppunct}\relax
\EndOfBibitem
\bibitem[Kresse and Furthm{\"u}ller(1996)Kresse, and
  Furthm{\"u}ller]{1996Efficient}
Kresse,~G.~G.; Furthm{\"u}ller,~J. Efficient Iterative Schemes for Ab Initio
  Total-energy Calculations Using a Plane-wave Basis Set. \emph{Phys. Rev. B}
  \textbf{1996}, \emph{54}, 11169\relax
\mciteBstWouldAddEndPuncttrue
\mciteSetBstMidEndSepPunct{\mcitedefaultmidpunct}
{\mcitedefaultendpunct}{\mcitedefaultseppunct}\relax
\EndOfBibitem
\bibitem[Bl\"ochl(1994)]{1994PAW}
Bl\"ochl,~P.~E. Projector Augmented-wave Method. \emph{Phys. Rev. B}
  \textbf{1994}, \emph{50}, 17953--17979\relax
\mciteBstWouldAddEndPuncttrue
\mciteSetBstMidEndSepPunct{\mcitedefaultmidpunct}
{\mcitedefaultendpunct}{\mcitedefaultseppunct}\relax
\EndOfBibitem
\bibitem[Togo and Tanaka(2015)Togo, and Tanaka]{2015phonopy}
Togo,~A.; Tanaka,~I. First Principles Phonon Calculations in Materials Science.
  \emph{Scripta. Mater.} \textbf{2015}, \emph{108}, 1--5\relax
\mciteBstWouldAddEndPuncttrue
\mciteSetBstMidEndSepPunct{\mcitedefaultmidpunct}
{\mcitedefaultendpunct}{\mcitedefaultseppunct}\relax
\EndOfBibitem
\bibitem[Chernatynskiy and Phillpot(2010)Chernatynskiy, and Phillpot]{2010BTE}
Chernatynskiy,~A.; Phillpot,~S.~R. Evaluation of Computational Techniques for
  Solving the Boltzmann Transport Equation for Lattice Thermal Conductivity
  Calculations. \emph{Phys. Rev. B} \textbf{2010}, \emph{82}, 1456--1461\relax
\mciteBstWouldAddEndPuncttrue
\mciteSetBstMidEndSepPunct{\mcitedefaultmidpunct}
{\mcitedefaultendpunct}{\mcitedefaultseppunct}\relax
\EndOfBibitem
\bibitem[Li \latin{et~al.}(2014)Li, Carrete, A.~Katcho, and
  Mingo]{2014ShengBTE}
Li,~W.; Carrete,~J.; A.~Katcho,~N.; Mingo,~N. ShengBTE: A Solver of the
  Boltzmann Transport Equation for Phonons. \emph{Comput. Phys. Commun.}
  \textbf{2014}, \emph{185}, 1747--1758\relax
\mciteBstWouldAddEndPuncttrue
\mciteSetBstMidEndSepPunct{\mcitedefaultmidpunct}
{\mcitedefaultendpunct}{\mcitedefaultseppunct}\relax
\EndOfBibitem
\bibitem[Ward \latin{et~al.}(2009)Ward, Broido, Stewart, and
  Deinzer]{2009-Diamond-P3}
Ward,~A.; Broido,~D.~A.; Stewart,~D.~A.; Deinzer,~G. Ab Initio Theory of the
  Lattice Thermal Conductivity in Diamond. \emph{Phys. Rev. B} \textbf{2009},
  \emph{80}, 125203\relax
\mciteBstWouldAddEndPuncttrue
\mciteSetBstMidEndSepPunct{\mcitedefaultmidpunct}
{\mcitedefaultendpunct}{\mcitedefaultseppunct}\relax
\EndOfBibitem
\bibitem[Zhou \latin{et~al.}(2018)Zhou, Fan, Qin, Yang, Ouyang, and
  Hu]{2018-MD-BTE-NEGF-ACSomega}
Zhou,~Y.; Fan,~Z.; Qin,~G.; Yang,~J.; Ouyang,~T.; Hu,~M. {Methodology
  Perspective of Computing Thermal Transport in Low-dimensional Materials and
  Nanostructures : The Old and the New}. \emph{ACS Omega} \textbf{2018},
  \emph{3}, 3278--3284\relax
\mciteBstWouldAddEndPuncttrue
\mciteSetBstMidEndSepPunct{\mcitedefaultmidpunct}
{\mcitedefaultendpunct}{\mcitedefaultseppunct}\relax
\EndOfBibitem
\bibitem[Asen-Palmer \latin{et~al.}(1997)Asen-Palmer, Bartkowski, Gmelin,
  Cardona, Zhernov, Inyushkin, Taldenkov, Ozhogin, Itoh, and
  Haller]{1997-Ge-debyeCallaway-PRB}
Asen-Palmer,~M.; Bartkowski,~K.; Gmelin,~E.; Cardona,~M.; Zhernov,~A.~P.;
  Inyushkin,~A.~V.; Taldenkov,~A.; Ozhogin,~V.~I.; Itoh,~K.~M.; Haller,~E.~E.
  Thermal Conductivity of Germanium Crystals with Different Isotopic
  Compositions. \emph{Phys. Rev. B} \textbf{1997}, \emph{56}, 9431--9447\relax
\mciteBstWouldAddEndPuncttrue
\mciteSetBstMidEndSepPunct{\mcitedefaultmidpunct}
{\mcitedefaultendpunct}{\mcitedefaultseppunct}\relax
\EndOfBibitem
\bibitem[Zhang(2016)]{2016-debye-yang}
Zhang,~Y. First-principles {Debye–Callaway} Approach to Lattice Thermal
  Conductivity. \emph{J. Materiomics} \textbf{2016}, \emph{2}, 237--247\relax
\mciteBstWouldAddEndPuncttrue
\mciteSetBstMidEndSepPunct{\mcitedefaultmidpunct}
{\mcitedefaultendpunct}{\mcitedefaultseppunct}\relax
\EndOfBibitem
\end{mcitethebibliography}

\newpage
\LARGE{\textbf{Supporting information\\ High and anomalous thermal conductivity in monolayer MSi$_2$Z$_4$ semiconductors}}

\newpage
\begin{figure*}[!t]
\centering
\includegraphics[width=16cm]{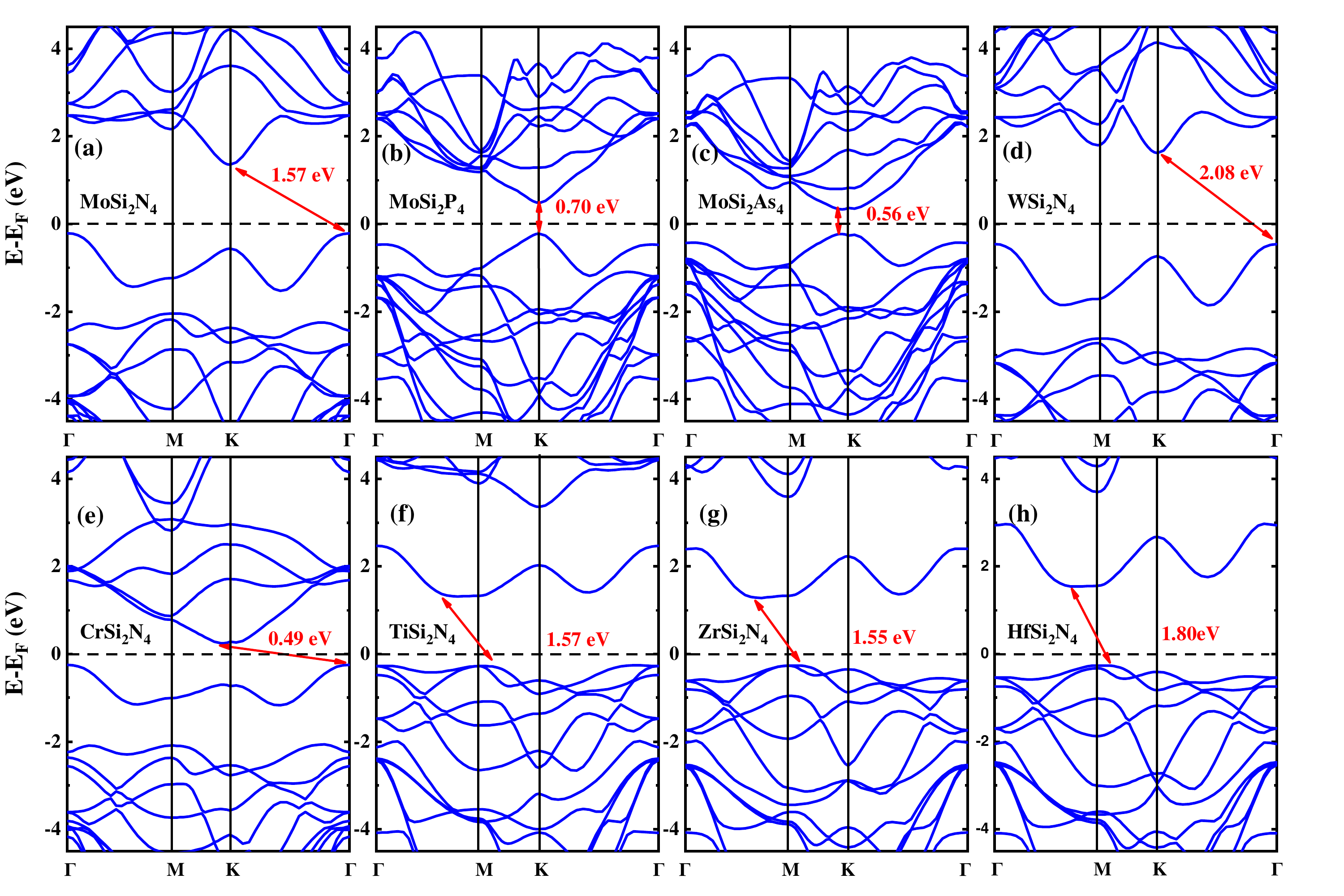}
\caption*{Figure~S1.~Electronic band structures of MSi$_2$Z$_4$: (a) MoSi$_2$N$_4$, (b) MoSi$_2$P$_4$, (c) MoSi$_2$As$_4$, (d) WSi$_2$N$_4$, (e) CrSi$_2$N$_4$, (f) TiSi$_2$N$_4$, (g) ZrSi$_2$N$_4$ and (h) HfSi$_2$N$_4$.}
\label{figS1}
\end{figure*}

\newpage
\begin{table*}[htbp]
\centering
\caption*{Table~S1.~Band gaps of 2D MSi$_2$Z$_4$ by GGA and HSE, respectively.} 
\renewcommand\arraystretch{1.5} 
\setlength{\tabcolsep}{1.1mm}
\begin{tabular}{ccccccccc}
\toprule
 &MoSi$_2$N$_4$ &MoSi$_2$P$_4$ &MoSi$_2$As$_4$ &WSi$_2$N$_4$ &CrSi$_2$N$_4$ &TiSi$_2$N$_4$ &ZrSi$_2$N$_4$ &HfSi$_2$N$_4$ \\ 
\midrule
GGA~(eV) &1.56 &0.70 &0.56 &2.08 &0.49 &1.57 &1.55 &1.80 \\
HSE~(eV) &2.40 &1.11 &0.98 &2.64 &1.11 &2.50 &2.41 &2.70\\
\bottomrule
\end{tabular}
\label{tableS1}
\end{table*}

\begin{table*}[htbp]
\centering
\caption*{Table~S2.~Born effective charges ($Z^*$) of each nonequivalent atom (M, Si, Z1, Z2) and dielectric constants ($\varepsilon$) of 2D MSi$_2$Z$_4$.} 
\renewcommand\arraystretch{1.5} 
\setlength{\tabcolsep}{3mm}
\begin{tabular}{ccccccc}
\toprule
  &component &$Z^*$~(M) &$Z^*$~(Si) &$Z^*$~(Z1) &$Z^*$~(Z2) &$\varepsilon$\\ 
\midrule
MoSi$_2$N$_4$ &xx &0.88 &3.30 &-0.84 &-2.90 &5.14\\ 
              &yy &0.88 &3.30 &-0.85 &-2.90 &5.14\\
              &zz &-0.02 &1.03 & -0.48 &-0.53 &1.65\\
MoSi$_2$P$_4$ &xx &-3.74 &1.81 &1.84 &-1.78 &10.82\\ 
              &yy &-3.74 &1.81 &1.84 &-1.78 &10.82\\
              &zz &-0.59 &0.28 &0.28 &-0.11 &2.27\\
MoSi$_2$As$_4$ &xx &-4.15 &1.49 &2.12 &-1.53 &12.76\\ 
              &yy &-4.15 &1.49 &2.12 &-1.53 &12.76\\
              &zz &-0.58 &0.18 &0.17 &-0.06 &2.49\\
WSi$_2$N$_4$ &xx &1.39 &3.31 &-1.16 &-2.85 &4.66\\ 
              &yy &1.39 &3.31 &-1.16 &-2.85 &4.66\\
              &zz &-0.02 &0.97 &-0.64 &-0.33 &1.64\\
CrSi$_2$N$_4$ &xx &-0.58 &3.21 &-0.03 &-2.89 &6.59\\ 
              &yy &-0.58 &3.21 &-0.03 &-2.89 &6.59\\
              &zz &-0.10 &0.91 &-0.38 &-0.47 &1.64\\
TiSi$_2$N$_4$ &xx &4.77 &3.29 &-2.99 &-2.68 &4.17\\ 
              &yy &4.77 &3.29 &-2.99 &-2.68 &4.17\\
              &zz &0.39 &0.83 &-0.74 &-0.28 &1.67\\ 
ZrSi$_2$N$_4$ &xx &5.02 &3.33 &-3.16 &-2.68 &3.96\\ 
              &yy &5.02 &3.33 &-3.16 &-2.68 &3.96\\
              &zz &0.56 &0.88 &-0.85 &-0.30 &1.70\\
HfSi$_2$N$_4$ &xx &4.63 &3.34 &-2.97 &-2.69 &3.61\\ 
              &yy &4.63 &3.34 &-2.97 &-2.68 &3.61\\
              &zz &0.60 &0.89 &-0.91 &-0.28 &1.68\\
\bottomrule
\end{tabular}
\label{tableS2}
\end{table*}

\begin{figure*}[!t]
\centering
\includegraphics[width=16cm]{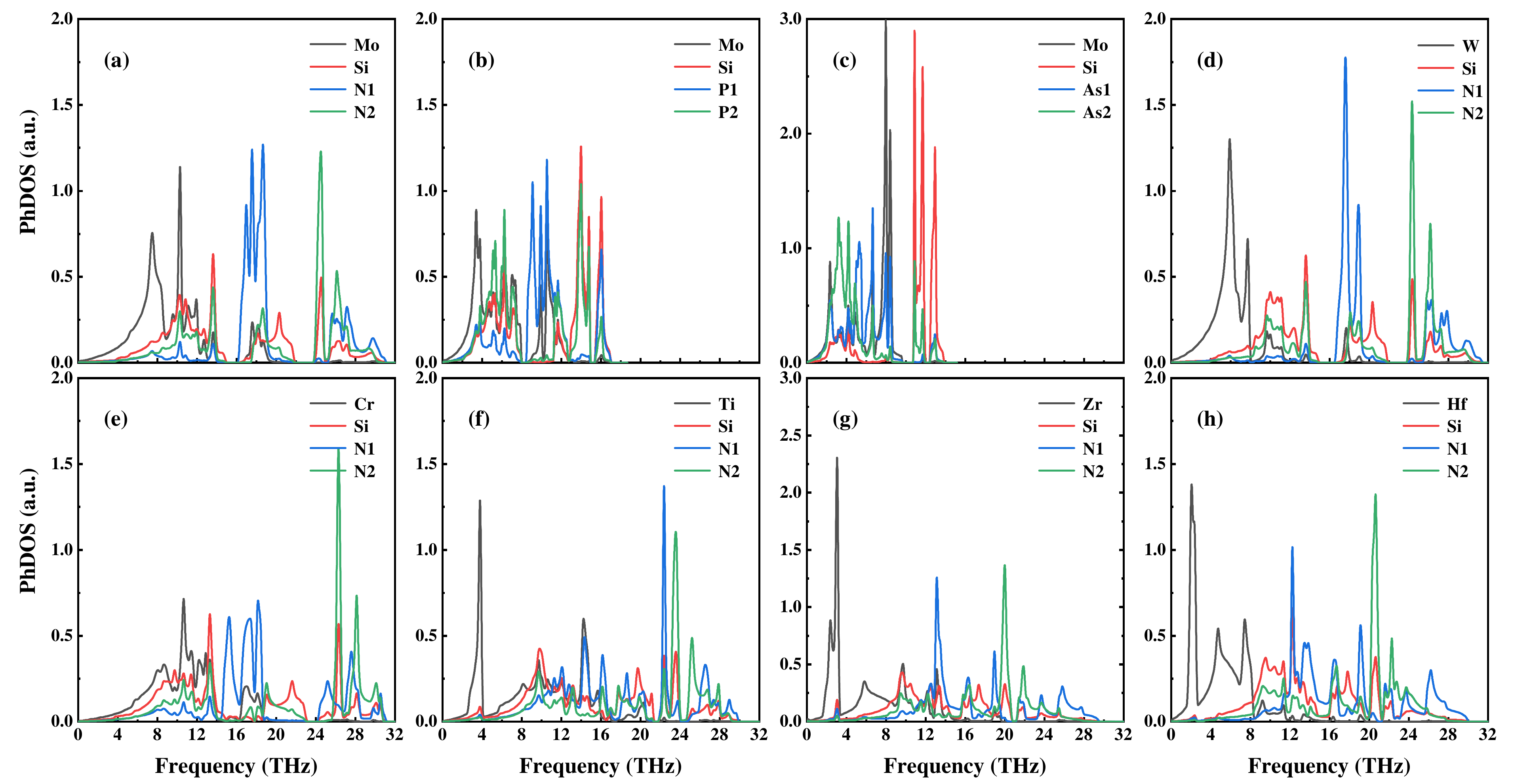}
\caption*{Figure~S2.~Phonon density of states of MSi$_2$Z$_4$: (a) MoSi$_2$N$_4$, (b) MoSi$_2$P$_4$, (c) MoSi$_2$As$_4$, (d) WSi$_2$N$_4$, (e) CrSi$_2$N$_4$, (f) TiSi$_2$N$_4$, (g) ZrSi$_2$N$_4$ and (h) HfSi$_2$N$_4$.}
\label{figS2}
\end{figure*}

\begin{figure*}[!t]
\centering
\includegraphics[width=8.6cm]{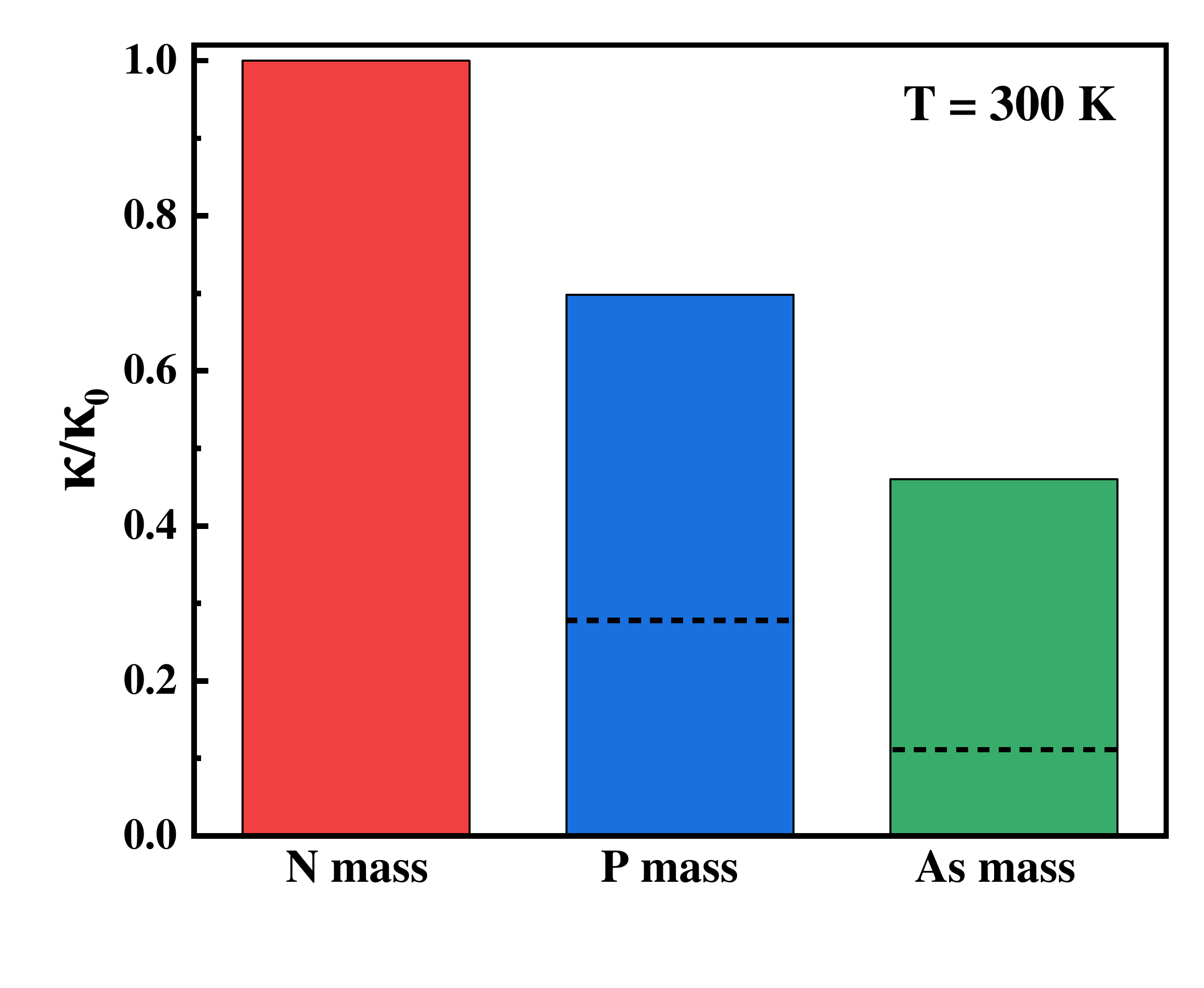}
\caption*{Figure~S3.~The mass effect of Z atom on the thermal conductivity of MoSi$_2$Z$_4$ by only changing the mass of N atom to P or As atom in MoSi$_2$N$_4$. $\kappa_0$ and $\kappa$ are the thermal conductivity of intrinsic and mass substituted MoSi$_2$N$_4$, respectively. Dash lines in blue (green) stripe represents the ratio between intrinsic $\kappa$ of MoSi$_2$P$_4$ (MoSi$_2$As$_4$) and $\kappa$ of MoSi$_2$N$_4$.}
\label{figS3}
\end{figure*}

\begin{figure*}[!t]
\centering
\includegraphics[width=16cm]{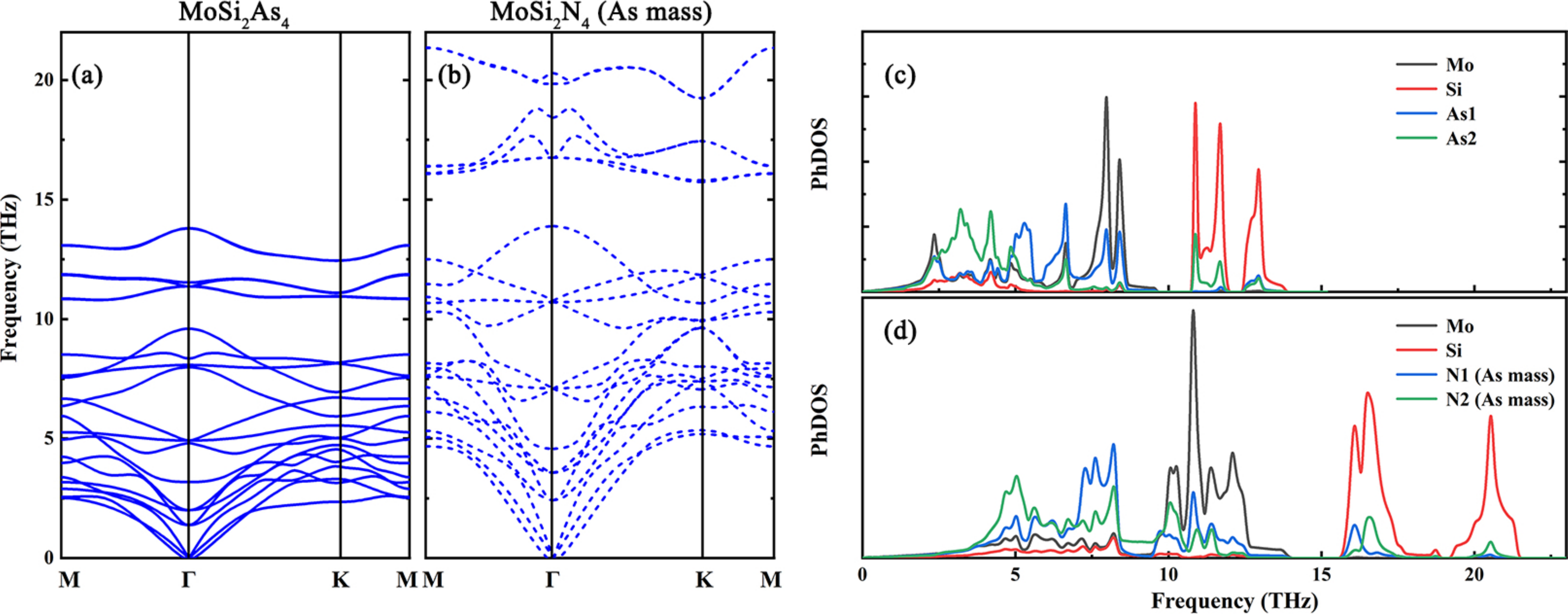}
\caption*{Figure~S4.~The effect of bond strength induced by Z atom on the phonon dispersion and phonon density of states of MoSi$_2$Z$_4$. For ensuring mass-independent, the mass of N atom is set as the mass of As atom.  (a) and (b) are the phonon dispersion spectrum of MoSi$_2$As$_4$ and MoSi$_2$N$_4$ with As mass, (c) and (d) are the corresponding phonon density of states.
}
\label{figS4}
\end{figure*}

\begin{figure*}[!t]
\centering
\includegraphics[width=16cm]{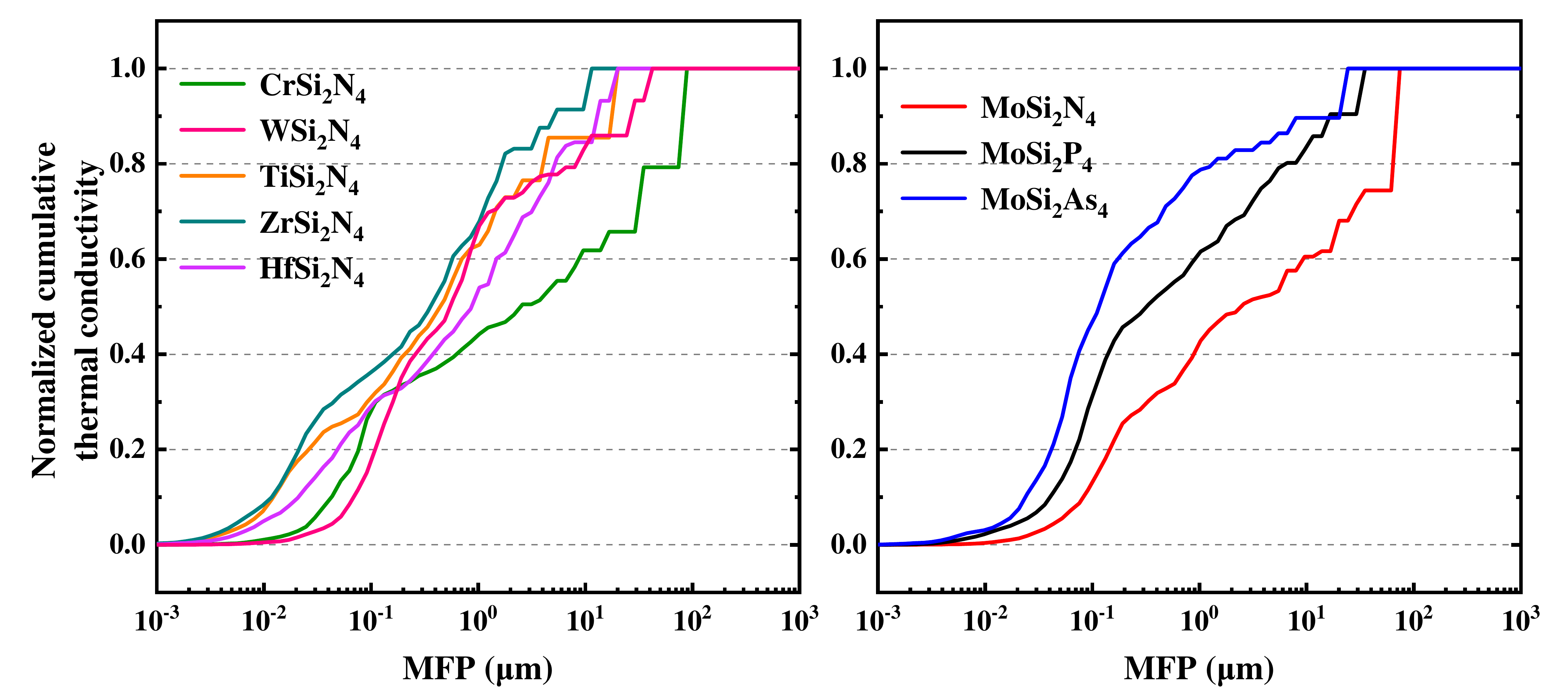}
\caption*{Figure~S5.~The normalized cumulative thermal conductivity as the function of mean free path (MFP) of CrSi$_2$N$_4$, WSi$_2$N$_4$,  TiSi$_2$N$_4$, ZrSi$_2$N$_4$, HfSi$_2$N$_4$, MoSi$_2$N$_4$, MoSi$_2$P$_4$ and MoSi$_2$As$_4$.
}
\label{figS5}
\end{figure*}

\begin{figure*}[!t]
\centering
\includegraphics[width=12cm]{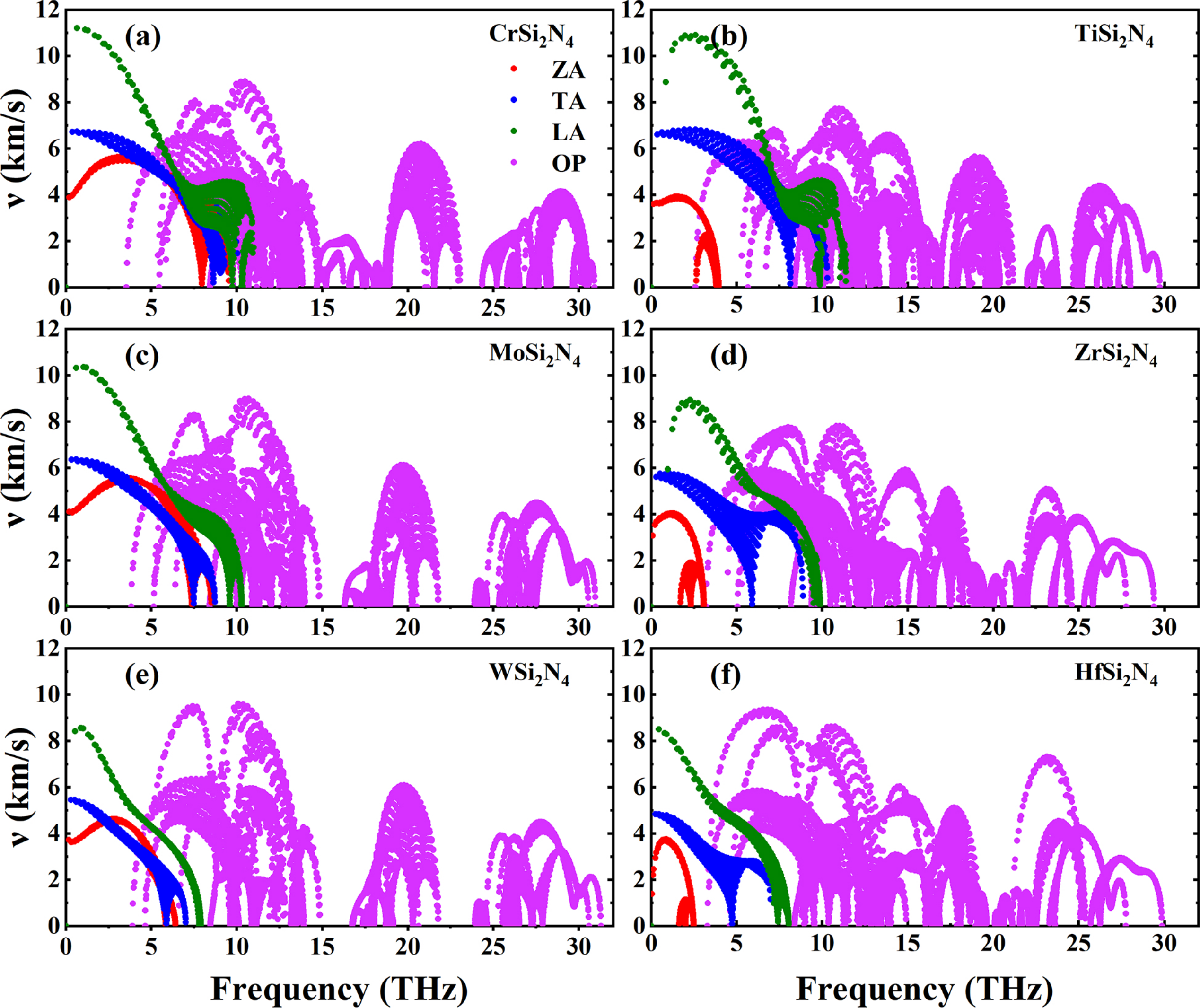}
\caption*{Figure~S6.~Phonon group velocity of (a) CrSi$_2$N$_4$, (b) TiSi$_2$N$_4$, (c) MoSi$_2$N$_4$, (d) ZrSi$_2$NZ$_4$ (e) WSi$_2$N$_4$ and (f) HfSi$_2$N$_4$.}
\label{figS6}
\end{figure*}

\begin{figure*}[!t]
\centering
\includegraphics[width=16cm]{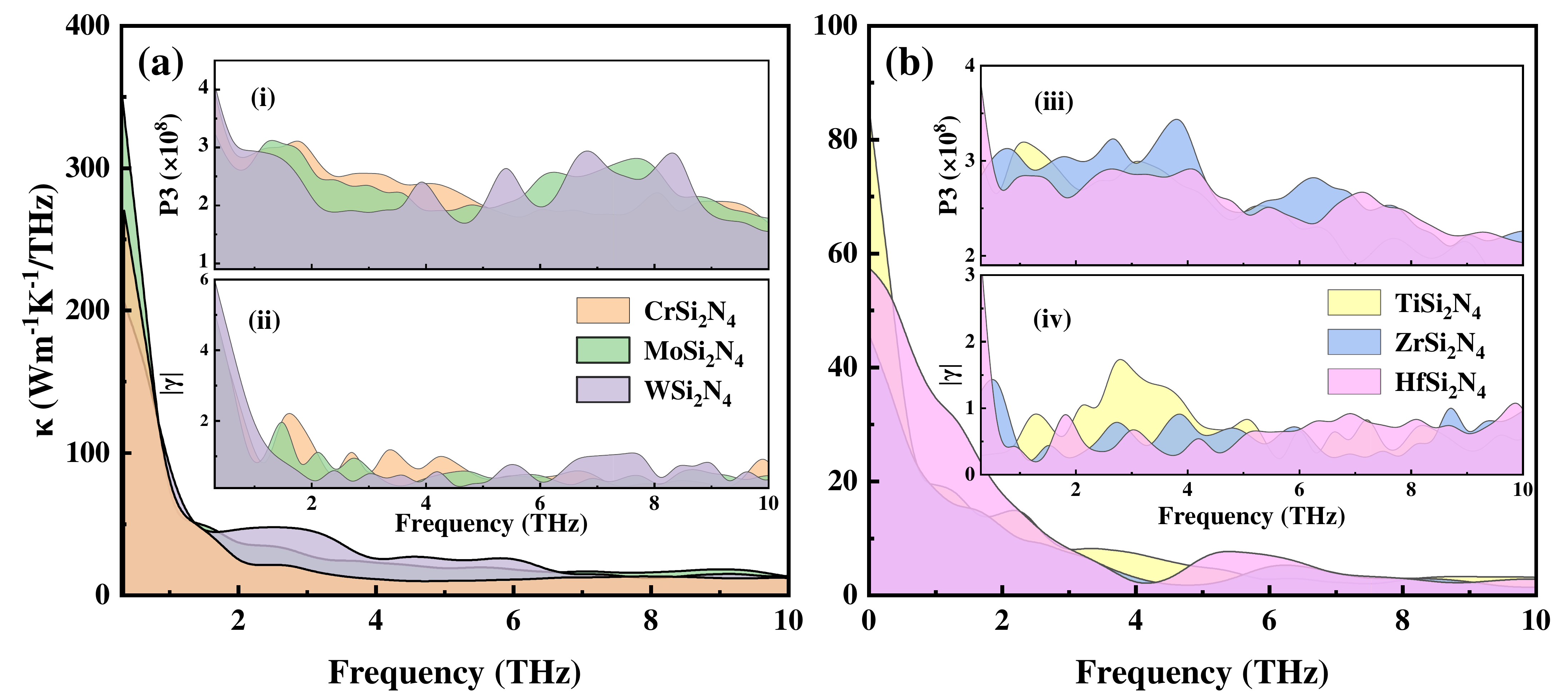}
\caption*{Figure~S7.~Room-temperature $\kappa$ as the function of frequency: (a) Group-\uppercase\expandafter{\romannumeral6}B and (b) Group-\uppercase\expandafter{\romannumeral4}B MSi$_2$N$_4$, the inserted of (\romannumeral1) and (\romannumeral3) are phase space (P3), (\romannumeral2) and (\romannumeral4) are absoluted Gr$\ddot{u}$neisen parameter ($|\gamma|$).
}
\label{figS7}
\end{figure*}

\begin{figure*}[!t]
\centering
\includegraphics[width=16cm]{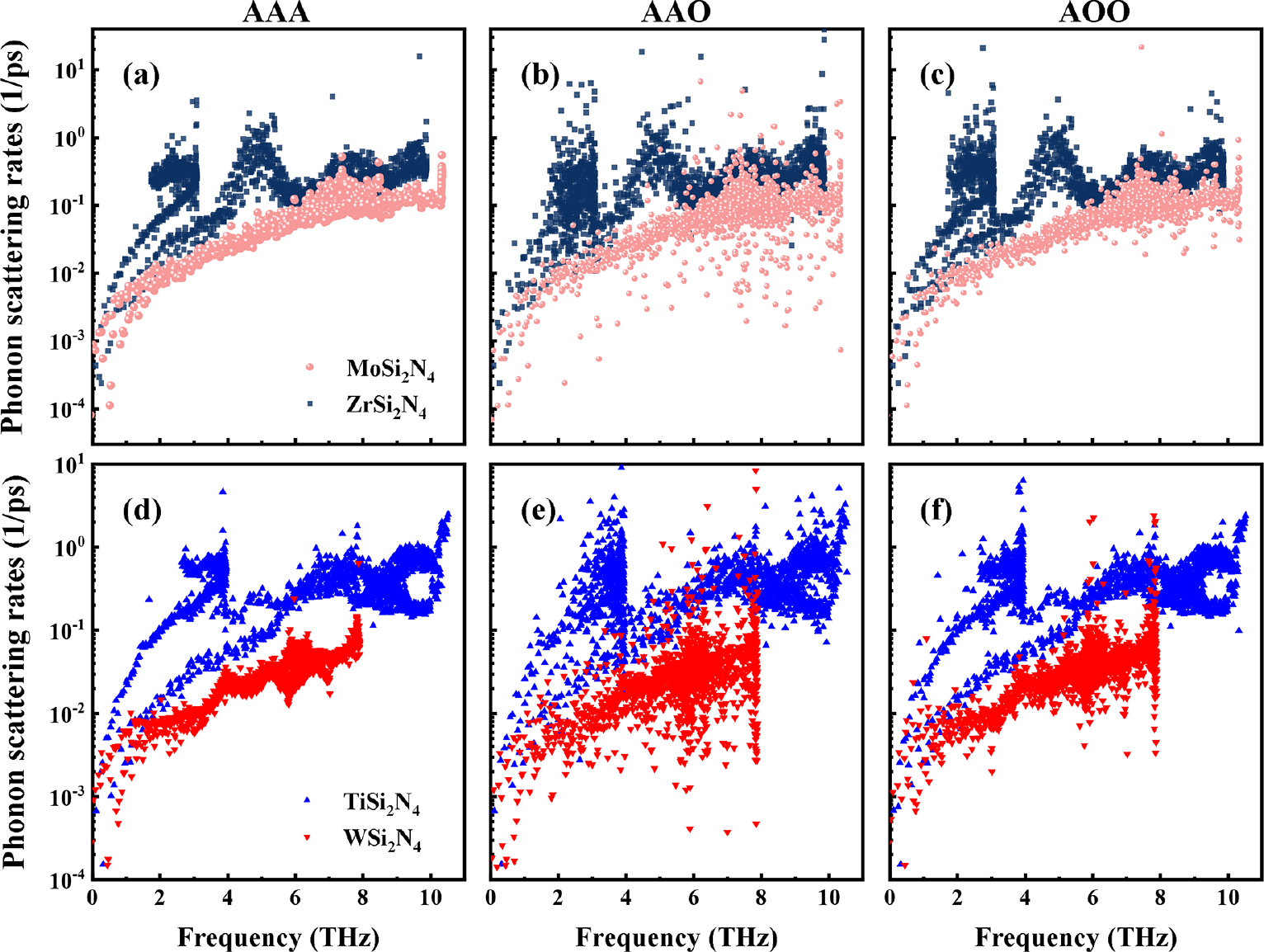}
\caption*{Figure~S8.~The comparison of phonon scattering rates for particular interacting channels (AAA, AAO and AOO) between (a)--(c) MoSi$_2$N$_4$ and ZrSi$_2$N$_4$, and between (d)--(f) TiSi$_2$N$_4$ and WSi$_2$N$_4$.
}
\label{figS8}
\end{figure*}

\begin{figure*}[!t]
\centering
\includegraphics[width=16cm]{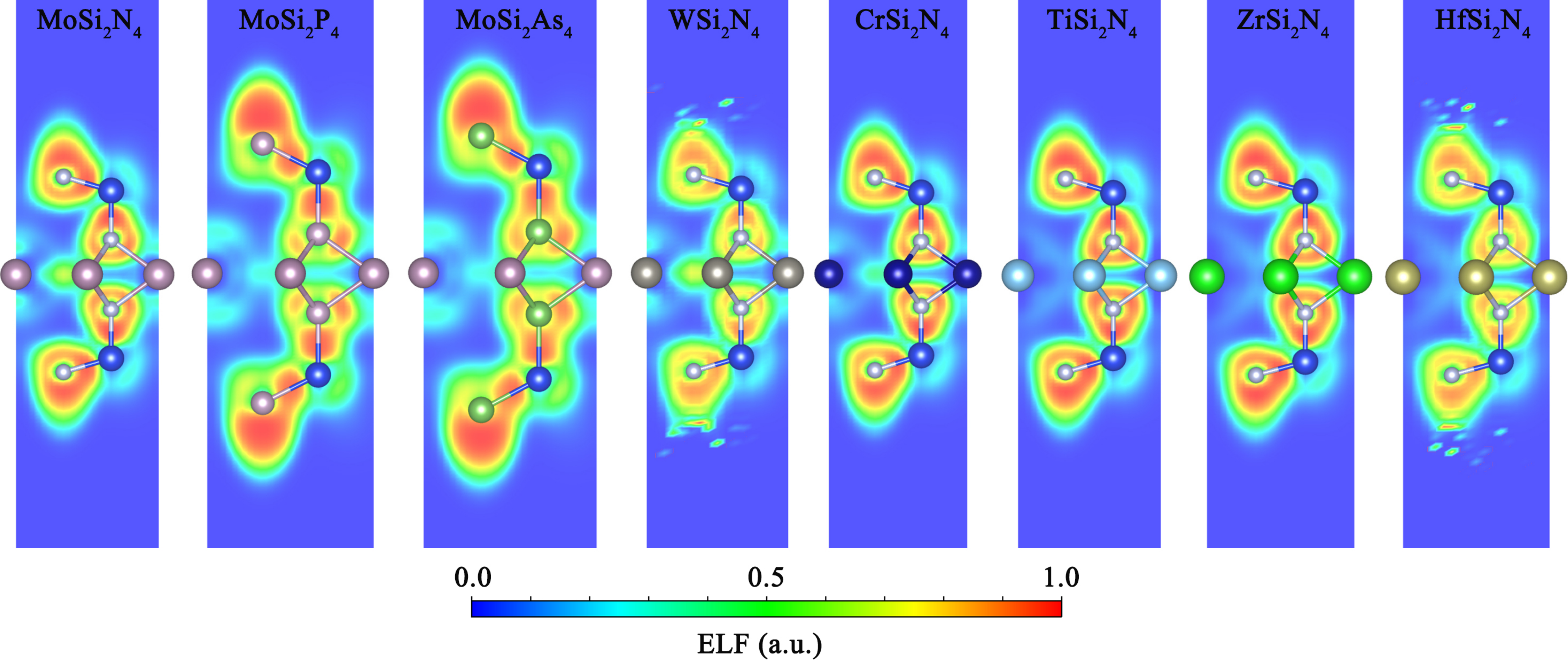}
\caption*{Figure~S9.~Electron localization function (ELF) distribution of MSi$_2$Z$_4$ in the (110) plane.}
\label{figS9}
\end{figure*}

\begin{figure*}[!t]
\centering
\includegraphics[width=8.6cm]{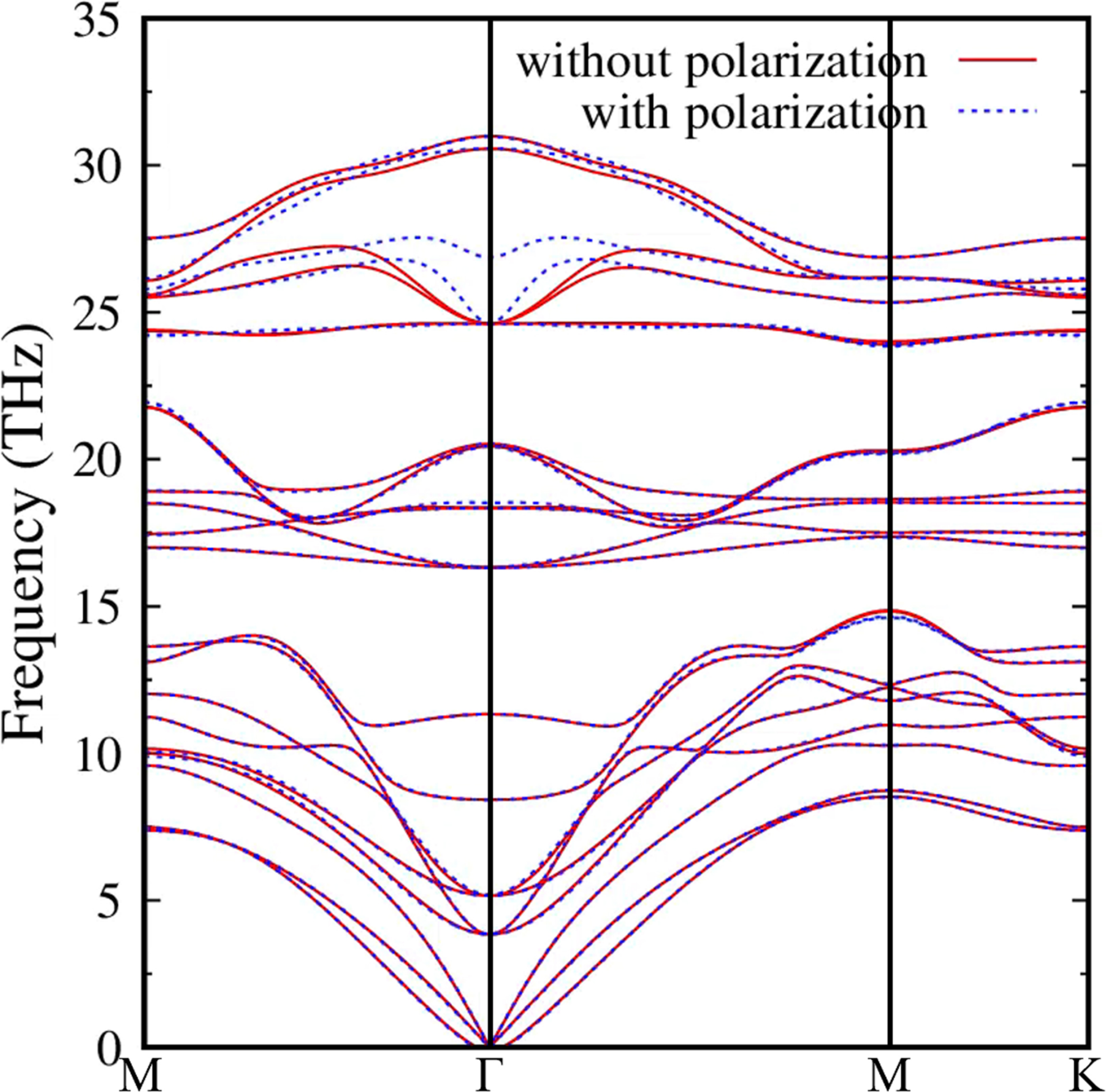}
\caption*{Figure~S10.~Phonon dispersion spectrum of MoSi$_2$N$_4$ with (blue dotted line) or without (red solid line) polarization effect.}
\label{figS10}
\end{figure*}

\begin{figure*}[!t]
\centering
\includegraphics[width=8.6cm]{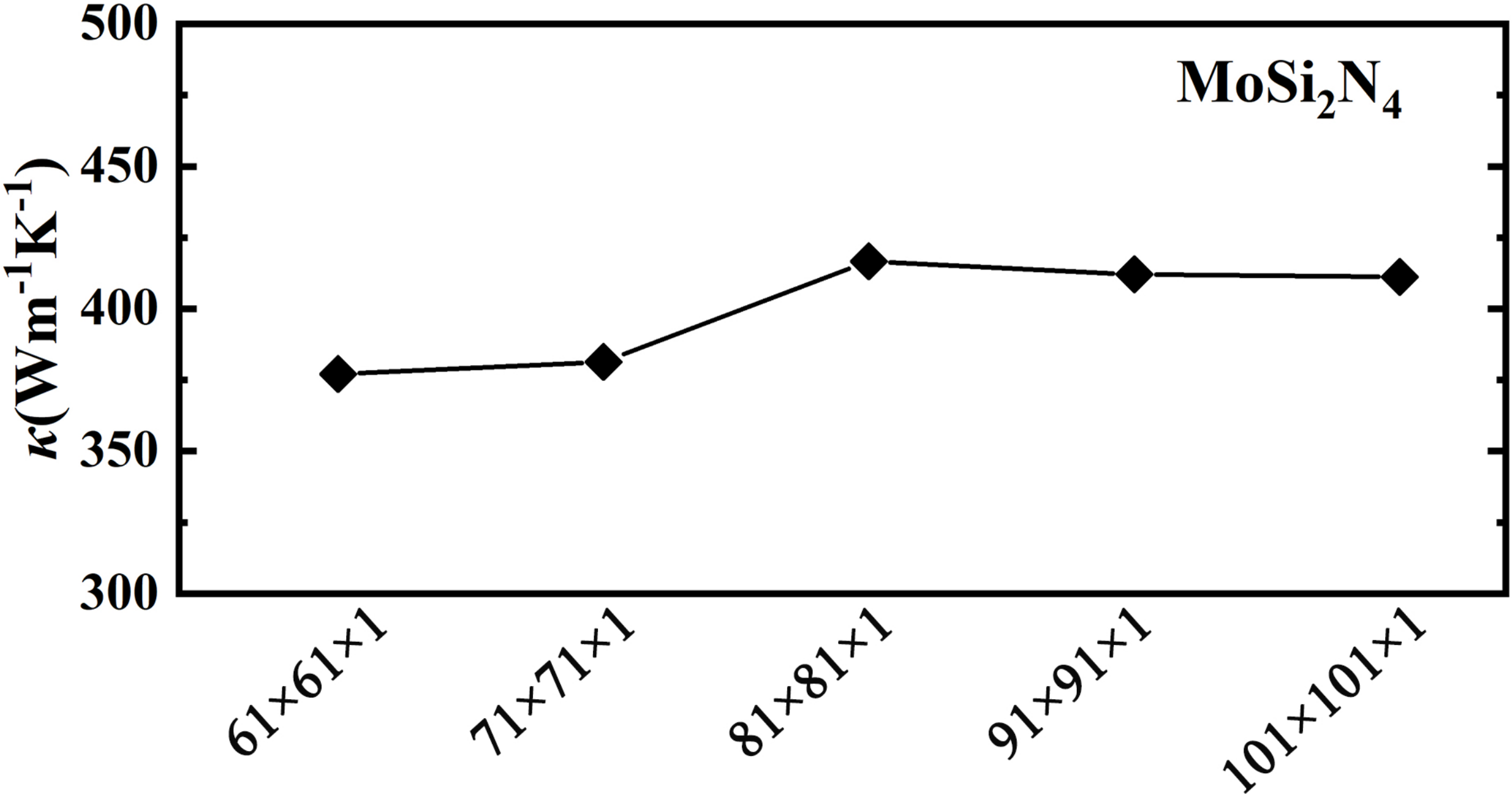}
\caption*{Figure~S11.~Lattice thermal conductivity ($\kappa$) of MoSi$_2$N$_4$ as function of grid density from 61$\times$61$\times$1 to 101$\times$101$\times$1}
\label{figS11}
\end{figure*}

\end{document}